Morphological classification and structural parameters for early-type galaxies in the Coma cluster.[*]

S. Andreon[1,**], E. Davoust[1], R. Michard[2], J.-L. Nieto[1,†], P. Poulain[1]

[1] CNRS–URA 285, Observatoire Midi-Pyrénées, 14 Avenue E. Belin, F-31400 Toulouse, France

[2] O.C.A., Observatoire de Nice, B.P. 229, F-06304 Nice Cedex 4, France

**Abstract.** – We present the results of an isophotal shape analysis of three samples of galaxies in the Coma cluster. Quantitative morphology, together with structural and photometric parameters, is given for each galaxy. Special emphasis has been placed on the detailed classification of early-type galaxies.

The three samples are: i) a sample of 97 early-type galaxies brighter than $m_B = 17.00$ falling within one degree from the center of the Coma cluster; these galaxies were observed with CCD cameras, mostly in good to excellent resolution conditions; ii) a magnitude complete sample of 107 galaxies of all morphological types down to $m_B = 17.00$ falling in a circular region of 50 arcmin diameter, slightly offcentered to the North-West of the cluster center; the images for this and the next sample come from digitized photographic plates; iii) a complete comparison sample of 26 galaxies of all morphological types down to $m_R = 16.05$ (or $m_B \simeq 17.5$), also in a region of 50 arcmin diameter, but centered 2.6 degrees West of the cluster center.

The reliability of our morphological classifications and structural parameters of galaxies, down to the adopted magnitude limits, is assessed by comparing the results on those galaxies for which we had images taken with different instrumentation and/or seeing conditions, and by comparing our results with similar data from other observers.

1. Introduction

Much effort has been devoted in the past decade to investigate the structure and dynamics of early-type galaxies, and, as a consequence, our understanding of their true nature has certainly improved. The "dichotomy" of ellipticals, which fall roughly into two classes, "disky" ellipticals with pointed isophotes, which are rotationally supported, and boxy/irregular ellipticals, probably triaxial systems, is now well established (Bender 1990; Nieto et al. 1994), and the continuity between the former and lenticulars is also becoming evident (Michard, 1994).

This recent progress has essentially been possible thanks to quantitative morphological classification, based on the analysis of isophote shapes and of luminosity profiles of large samples of galaxies observed with CCD detectors. We contributed to this effort by two series of papers, one series presenting the isophote shapes of selected samples of ellipticals (Nieto et al. 1991, Poulain et al. 1992), and their analysis (Nieto et al. 1994, Andreon 1994), and the other presenting the quantitative morphology of E-S0 galaxies (Michard & Marchal 1993, 1994a,b) and their analysis (Michard 1994).

We may now ask if the observable morphological properties of early-type galaxies show significant changes with their environment, and for this purpose, the study of a rich and dense cluster like Coma is in order. The properties of its population of E-S0 galaxies may be compared with those of the Local Supercluster (Michard 1994) and of the Perseus (Andreon 1994) and Virgo clusters.

In this paper, we present the results of an isophotal shape analysis of three samples of galaxies in the Coma cluster. The three samples, together with their observations, are fully described in Sect. 2. The techniques of analysis followed in the present study are summarized in Sect. 3. The results are given in Sect. 4 in the form of tables (Tables 2, 3 and 4), one for each sample; the tables include global photometric and geometrical parameters as well as detailed morphological information, but not the photometric and geometrical profiles as a function of radius. The latter will be made available in electronic form at the CDS in Strasbourg. Notes on individual galaxies follow, in Sect. 5. A preliminary discussion is given in Sect. 6, including a comparison with previous isophotal analyses of galaxies in the Coma cluster.

These results will be used in subsequent papers to study properties of galaxies in different environments, the morphological segregation of galaxies in clusters and the continuity between ellipticals and lenticulars, among other projects (see Andreon 1994 and Michard 1994, 1995 for further details).

## 2. The three samples and their observations

The three samples were selected with the aim of covering somewhat different "environments", i.e. distances to the cluster center. We decided to use both photographic and CCD data; the latter are expected to give better accuracy on all kind of measurements, but it is much easier to obtain complete samples from the former.

### 2.1 The CCD sample

This sample of galaxies was taken from the catalogue of Godwin et al. (1983; hereafter GMP), which lists all galaxies down to magnitude $m_B = 20.0$ in a 2.63 degree square area, centered on the Coma cluster. We selected all the galaxies classified early-type in the literature down to magnitude $m_B = 17.0$, within one degree from the center. This limiting magnitude was adopted in order to have a manageable number of galaxies to observe. This sample might not be complete in magnitude (in the sense that we could have missed true early-type objects), because there is no catalogue of galaxies complete to our magnitude limit *and* with morphological classifications. Figure 1 shows the spatial distribution of the galaxies in the sample.

The galaxies of this sample were observed with the 2-meter Télescope Bernard Lyot (hereafter TBL) at Pic du Midi Observatory in January (1) and March 1990 (2), March 1991 (3), January 1992 (4), February (6), March (7) and May 1993 (8), and March 1994 (9); the number of the run is given in parentheses. All galaxies were observed through an R filter; V images for 12 galaxies, obtained in 1990, were only used to check the stability of the results of our isophotal analysis. The seeing ranged from 0.8" to 3.80" (FWHM), with a mean at 1.48". The CCD was an RCA 512×320 (pixel size 0.324") in the early runs, and a THX 1024×1024 (pixel size between 0.21" and 0.48", depending on focal reducer and binning) in the others. We also had a run (run 5) at the CFH 3.6-meter telescope (hereafter CFHT) in Hawaii in April 1992, with excellent seeing (between 0.35" and 0.80"). We used the High Resolution Camera (HRCam) with the RCA4 1024×640 camera (pixel size 0.108") and an R filter. All CCD images were bias subtracted and flat-fielded in the usual way. The charge skimming present in the CFH frames was removed in the way presented by Andreon (1993).

The observing log is described in Table 1. (1) : abriged GMP or NGC number; (2) : exposure time in seconds; (3) : filter; (4) observing run (see above); (5) : seeing (FWHM) in arcsec; (6) and (7) : sky brightness and level of last measured isophote, in mag arcsec$^{-2}$; (8) : notes. In run 7, we used a standard magnitude constant for the whole run. In run 9, we derived a standard magnitude constant for each night. This table will be published in electronic form only.

### 2.2 The KP1608 & FW samples

Two magnitude-complete samples of galaxies in Coma, the KP1608 sample and the FW sample, were selected from photographic plate material.

Two photographic plates were obtained from the collection of plates of the Mayall 4-meter telescope, among fifteen covering the Coma cluster, described in Strom & Strom (1978; their Table II). One plate (KP1608) covers the central region; it is centered 7.5 arcmin North and 19.4 arcmin West of NGC 4889. The other plate (KP2225) covers a region near the western border of the cluster; it is centered 156 arcmin West of NGC 4889. The field of view of the two plates is 54 arcmin in diameter, but the used field was restricted to 50 arcmin in diameter. The exposure times were 40 and 60 minutes respectively, a red (RG-610) filter was used. The seeing was about 1.9" (FWHM) for KP1608 and slightly better (1.8") for KP2225. Figure 2 and 3 show the spatial distribution of galaxies in the KP1608 and FW regions respectively. A reproduction of plate KP2225 can be found in Strom & Strom (1978; their Fig. 13).

Seeing is a problem when classifying galaxies. Since seeing affects smaller galaxies first, we chose the limiting magnitude of each sample in such a way that we would just start to have difficulties in classifying the faintest and smallest galaxies of the sample. In other words, the limiting magnitude is slightly above the limit at which the determination of morphological types becomes uncertain. For the CCD sample, the limiting magnitude is generally well above that limit.

We selected all galaxies brighter than $m_B = 17.00$ on plate KP1608 (this is the KP1608 sample) and all galaxies brighter than $m_R = 16.05$ on plate KP2225 (the FW sample). No complete catalogue has yet been published for the FW region, so that, in this region, the galaxies were selected by visual inspection of the plate, and the limiting magnitude is in R. Because the adopted limiting magnitude is much brighter than the magnitude at which galaxies look like stars, the FW sample is also complete in magnitude. $m_R = 16.05$ corresponds roughly to $m_B = 17.5$.

The plates were digitized at the MAMA[1] with a 30 micron (= 0.557") aperture and step. This produces image files with pixel readings proportional to plate density. These were transformed into intensities, at a late stage of the isophotal analysis, by means of a contour to contour correspondence between plate-image output and CCD-image intensity, for a set of galaxies. For the calibration of plate KP1608, we used 10 images of the CCD sample. For the calibration of plate KP2225, we used CCD images of NGC 4692 and FW9, respectively taken at the 120cm telescope of Observatoire de Haute-Provence and at TBL. We checked that this technique produces good results in the extended range where the *sky corrected* density is proportional to the local surface brightness (hereafter SuBr or $\mu$) expressed in *magnitudes* per unit area. Uncertainties occur in the central parts of bright objects, due to differences in seeing between the plates and the calibrating CCD frames, and also to a possible saturation.

### 3. The photometric and isophotal analysis and the classification scheme

The photometric and isophotal profiles for all galaxies were obtained with a set of procedures running under the MIDAS image treatment software. The successive preliminary steps are : mapping of the sky background, inventory and elimination of all parasitic objects (cosmic rays, defects, galaxies and stars) close to the galaxy to be analysed, determination of the point spread function, determination of the magnitude zero point. The R magnitudes and surface brightnesses quoted in this paper refer to Cousin's R band, whereas the B and V magnitudes refer to the Johnson system.

The isophotal analysis itself follows the method of Carter (1978). The image is first smoothed by a method of "noise cheating enhancement", which does not affect the resolution in the bright parts of galaxy, but heavily smoothes the regions of low signal.

An analytic representation is then adjusted to each isophotal contour, involving a reference ellipse, and a Fourier series expansion of deviations from this ellipse in terms of the eccentric anomaly: we call here $e_i$ and $f_i$ the coefficients of the cosine and sine terms respectively, following Michard & Marchal 1993; these quantities are also called $a_i$ and $b_i$ in the literature, in particular in Nieto et al. (1991) and in Poulain et al. (1992). We refer to Michard & Marchal (1994a) for further details on the method of analysis.

One of the results of this analysis is the so-called "quantitative morphology", also described and examplified in the paper quoted above. We emphasize that the morphological classification proposed here is quantitative in the sense that it is based on the rational analysis of numerical data (photometric and geometrical profiles), rather than on the subjective judgment, based on galaxy images (and possibly on preconceived ideas of what they should look like), of a classical morphologist; the quantitative classification can thus be reproduced with a very high degree of accuracy.

A short summary is presented here for clarity:

1) the segregation between the E and S0 types is based upon the run of the SuBr along the major axis of the projected galaxy, plotted with an $r^{1/4}$ scale. The presence of a disk gives a characteristic bump above the linear profile which characterizes pure spheroids. This photometric signature of S0's should not be confused with other often observed deviations from the de Vaucouleurs law, both in giant ellipticals with the envelope enhanced above the extrapolated $r^{1/4}$ line, and in minor objects with evidence for a cut-off below this line. Note that this photometric criterion was used here to decide between the E and S0 cells of the morphological classification, while, in Michard & Marchal (1994a), it was only used in the cases where the classical morphological catalogues did not agree. Also note that it is identical, at least qualitatively, to the criterion used in classical morphology, when the observer looks for subtle changes of gradient in a galaxian image.

---

[1] MAMA (Machine Automatique à Mesurer pour l'Astronomie) is operated by CNRS/INSU

2) The E classified galaxies were subclassified into "disky" (or diE), "boxy" (or boE), and "undefined" (or unE). This is done from the profile of the $e_4$ coefficient. The unE are roundish galaxies where the $e_4$ values fluctuate around zero.

3) The S0's are subclassified into the SA0, SAB0 and SB0 families, based upon the run of the position angle (hereafter PA) of the isophotal major axis. For a galaxy observed with sufficient resolution, the PA of the bulge and disk are nearly the same, while the PA of the bar shows up in a limited range (except in cases of unfavorable projection). At the distance of the Coma cluster, it often happens that the contrast between bulge and bar is washed out by the seeing.

4) The results of our "quantitative morphology", presented below, also involve a classification of the *disk extent* and the *envelope geometry*, as introduced by Michard & Marchal (1993). This is based upon the run of the isophotal ellipticity and the $e_4$ coefficient, or more specifically upon the comparison of their value, at their maximum on the one hand, and in the galaxian envelope on the other. The classification cells for the envelope are "spheroidal halo" (or spH), "extended disk" (or exD), the intermediate case being "thick disk" (or thD). Disks may be termed emDi, if "embedded" in an envelope of the spH and thD classes, exDi if "extended" in an exD envelope. The intermediate case is miDi, when the disk appears "mixed" in a less flat envelope.

5) Finally, we give a few global parameters, such as the asymptotic magnitude, the effective radius, the semi-major axis at the effective isophote and the mean SuBr inside it, a typical axis ratio and $e_4$ parameter. These parameters do NOT constitute the quantitative morphology and, in fact, most of them are not used in the galaxy classification; they are just indicative, as galaxies with the same luminosity and geometrical profiles (thus the same classification) should have comparable values of these parameters (within a zero-point or scaling effect).

Most spiral galaxies were detected by a preliminary visual inspection of the images. Only in a few cases did the run of several of Carter's coefficients, implying large isophotal twists and asymmetries, prompt us to shift a disk galaxy into the S0/a or Sa cells. All the adopted morphological types were checked *a posteriori* by visual inspection of the galaxy images.

## 4. Results of the analysis

*4.1 Presentation of the data*

The data presented in Tables 2a, 2b, 3 and 4 include the usual photometric parameters in Cousin's R band, namely the asymptotic magnitude, the effective radius, the corresponding isophotal major axis, and the average SuBr inside the effective isophote. Geometrical parameters are given next, the minimum axis ratio (or alternatively its value at the effective isophote), the representative $e_4$ coefficient, the axis-ratio in the envelope, i.e. at the isophote $\mu_R = 24$ mag arcsec$^{-2}$, and a representative figure for the isophotal twist. Then comes a coded description, indicating the detection or absence of such components as a bar, a disk, a spiral pattern, and the classification of disks and envelopes.

Table 2a presents the results for the galaxies observed at TBL. It contains 97 galaxies brighter than $m_B = 17.00$, and falling within one degree from the cluster center. We also included 7 galaxies fainter than the limiting magnitude (RB 007, RB 039, RB 182, RB 183, RB 198, RB 261, RB 271), as well as one bright galaxy (NGC 4827) lying just outside the one degree region; they are listed at the end of Table 2a.

Table 2b presents the results for the galaxies observed at CFHT. It contains 17 galaxies brighter than $m_B = 17.00$, and falling within one degree from the cluster center, 13 of which were also observed at TBL. The asymptotic magnitudes of most bright galaxies of this subsample are uncertain, due to the small field of view of the CCD.

Table 3 presents the results for the KP1608 sample (107 galaxies), of which 35 are in common with the CCD

sample. We also included 6 galaxies fainter than the limiting magnitude; they are listed at the end of Table 3.

Table 4 presents the results for the FW sample. There are 26 galaxies brighter than $m_R = 16.05$ (or $m_B \simeq 17.5$) and 21 fainter galaxies, which are listed separately.

*4.2 Description of the table columns*

The 6 first columns concern the catalogue data, the others list the measured data.
(1) and (2) Number in the abridged and unabridged versions of the Goodwin, Metcalf and Peach (GMP) catalogue. (1950.0) coordinates in Table 4; these were taken from the APM digitized sky survey, or linearly interpolated from the RC3 (de Vaucouleurs et al. 1991) ones for galaxies missed by the APM survey (and by RC3).
(3) Number in Dressler's catalogue; empty in Table 4.
(4) Usual designation, such as NGC, IC, and RB (Rood & Baum 1967) numbers; in Table 4, FW and D from Strom & Strom (1978), anon = anonymous ($m_R \leq 16.05$), suppl = anonymous ($m_R > 16.05$), (see also Fig. 3).
(5) and (6) Differential coordinates, from the GMP catalogue; empty in Table 4.
(7) Asymptotic magnitude, in R (Cousin's system).
(8) Logarithm of the effective radius, in units of 0.1 arcmin.
(9) Logarithm of the semi major axis of the effective isophote, in units of 0.1 arcmin.
(10) Mean SuBr inside the effective isophote, in R mag arcmin$^{-2}$.
(11) Photometric evidence for a disk, coded as st (strong), cl (clear), ft (faint), or no (none).
(12) Typical axis ratio, either its minimum value, if clearly defined, or its value at the effective isophote otherwise.
(13) Typical $e_4$ parameter, either its extremum value, if clearly defined, or its value at the effective isophote otherwise. The estimates are in %. For the plate data, they were rounded to the nearest integer value.
(14) Location where the $e_4$ parameter was estimated, coded as ex (at its extremum), re (at the effective isophote), or co (if the value is the same at both locations). Note that the same indication applies also to the measurements of the axis ratio in column (12).
(15) Axis ratio in the envelope, i.e. at the isophote $\mu_R = 24$ mag arcsec$^{-2}$.
(16) Amplitude of isophotal twist in the range of reliable measurements, in degrees.
(17) Detection of a bar, coded as follows : bar (bar seen), bar? (bar suspected), -no (no bar seen).
(18) Detection and classification of a disk, coded as follows: emDi (embedded disk), miDi (mixed disk), exDi (extended disk), -?Di (detected but unclassified disk), -no- (no disk seen).
(19) Detection of a spiral pattern, coded as follows: spiP (spiral pattern seen), spiP? (spiral pattern suspected), -no- (no spiral pattern seen).
(20) Classification of an envelope, coded as follows: spH (spheroidal halo), thD (thick disk), exD (extended disk), pec (peculiar envelope), -?- (unclassified envelope).
(21) Our morphological classification, coded as follows: boE (boxy E), unE (undetermined E), diE (disky E), SA0, SAB0, SB0, Sa, etc., S... (spiral of unknown stage).
(22) A * refers to a note in Sect. 5 about specific features such as important dust pattern, ring or lens, low SuBr, $f_4$-asymmetry, etc., and about uncertainties of various sources.

*Nota Bene*: When a parameter has not been measured, or when a specific morphological component has not been studied, the relevant code is replaced by a dash.

## 5. Notes on individual galaxies

We give below qualitative remarks for galaxies which presented either peculiar morphological properties or practical problems for classification.

*5.1 CCD TBL subsample*

0238 (= NGC 4944): Curious disk structure, with two positive maxima of $e_4$, separated by an intermediate area with negative $e_4$, while $e_6$ remains large.

0334 (= NGC 4934): Dusty galaxy with asymmetric and non concentric isophotes, probably an S. Classified S in GMP and S? in RC3. Inner ring giving a sharp peak of $e_4 = 0.12$.

0421 (= NGC 4929): An envelope with strong twist, decreasing axis ratio, and disky isophotes is found.

0485: Inner ring and lens.

0504: Several shape coefficients differ from zero, which is evidence for irregular isophotes. Confirmed by visual inspection of the image.

0539: Inner ring and lens.

0591: There are uncertainties in the analysis of this roundish object. The profile follows the $r^{1/4}$ law, while a bar-like twist of the inner isophotes is observed.

0661 (= NGC 4923): There is evidence for an inner embedded disk in this roundish elliptical with a very large and regular isophotal twist.

0694: This galaxy shows a large spheroid, surrounded by a disky envelope well above the extrapolated SuBr of the "bulge".

0718 (= NGC 4919): Inner ring and lens.

0857: Marginal evidence for a low-contrast bar.

0867: Uncertain analysis, due to the presence of low surface brightness fluctuations which may or may not be intrinsic to the galaxy.

0876 (= NGC 4906): There is evidence for an inner embedded disk.

1035 (= RB 167): The galaxy has an inner boxy structure and a disky envelope. The boxiness is attributed to a bar, slightly tilted within the disk.

1039 (= NGC 4889): A large cD galaxy with a very extended envelope. The field of the CCD was not sufficient to extend the analysis to low SuBr.

1050 (= RB 077): Strong blend with NGC 4889. It was measured after subtraction of a spheroidal image of the cD galaxy.

1065 (= NGC 4886): Blend with NGC 4889.

1115 (= RB 155): A dominant bar is surrounded by a roundish envelope.

1198 (= RB 41): Companion to the West.

1204: Dust in the southern part of the disk of this North-South oriented galaxy. An S galaxy almost edge-on?

1214 (= NGC 4875): Marginal evidence for a bar in a lens.

1258 (= RB 026): The analysis is uncertain due to the halo of a very bright nearby star.

1277 (= RB 022): The large twist of the outer isophotes may be related to the halo of a very bright nearby star.

1300 (= RB 014): The large bulge looks like a bar along the major axis.

1378 (= NGC 4864): Uncertain analysis, due to a parasitic star 7 arcsec from the galaxian center.

1406 (= IC 3957): Our V frame would lead to the diE class, but this is contradicted by the R frame. A dust ring has been detected.

1414 (= IC 3955): This galaxy contains a small bulge at PA circa 70°, then a large dominant bar at PA = 45°, surrounded by a more roundish, strongly twisted envelope. S-shaped isophotes occur at the tips of the bar. It shows similarities with NGC 3414 (Michard 1994).

1507 (= NGC 4854): The bar is suspect, because the stellar images are elongated, but the KP1608 data confirm its presence.

1524: Irregular isophotes.

1646: This galaxy presents noticeable asymmetries, but its luminosity profile deviates very little from the $r^{1/4}$ law; doubtful classification.

1823 (= NGC 4842a): Marginal photometric evidence for a faint disk in this roundish galaxy, while $e_4$ remains zero.

1828 (= NGC 4841b): Difficult analysis because it is a roundish galaxy in the halo of NGC 4841a.

1831 (= NGC 4841a): The large twist of the outer isophotes is uncertain, due to the difficulty of separating NGC 4841a and b.

1834 (= NGFC 4840): The suspected bar is probably spurious (elongated stellar images).

1878 (= NGC 4839): The low average SuBr suggests that this galaxy is dominated by an extended disk. The galaxy is disky inside $r_e$ and boxy outside.

2074: Very large twist of the nearly circular isophotes.

2134 (= NGC 4816): This giant elliptical has an extended envelope well above the $r^{1/4}$ law relevant to the inner parts.

*5.2 CCD CFHT subsample*

0733 (= RB 129): Data from a frame kindly provided by C. Lobo (Institut d'Astrophysique de Paris).

1039 (= NGC 4889): Global photometry cannot be obtained from small field frames of this huge object.

1065 (= NGC 4886): The field was not sufficient to give reliable results for the integrated photometry. The total magnitude is underestimated.

1331 (= NGC 4865): This object shows a dust ring of circa 1" radius.

1883: This galaxy is a companion to a massive E, i.e. NGC 4839, so that its analysis refers to a limited radial extent. It is tempting to classify it among the small boE's "satellites", recognized as a specific group by Nieto and Bender (1989).

2109: This object is projected against the envelope of NGC 4816.

*5.3 KP1608 sample*

0884 (= IC 4040): Important dust pattern.

0969 (= RB 094): The bar is the dominant component.

0976 and 0980 (= NGC 4898b and a): A pair of overlapping early-type objects, with their centers about 6" apart.

1035 (= RB 167): The $e_4$ coefficient oscillates between an inner negative value, tentatively associated with a bar, and an outer positive one. The values of $e_4$ in Table 2a and 3 disagree only because they were not measured at the same isophote!

1039 (= NGC 4889): It is the eastern component of the famous pair of cD galaxies near the cluster center. It displays a very extended envelope.

1050 (= RB 077): It lies inside NGC4 889. A model of the giant galaxy was substracted from the frame, before measuring the companion.

1109 (= NGC 4883): This object displays a strongly twisted and asymmetric envelope. The morphological type is uncertain.

1115 (= RB 155): This object looks like a strong bar surrounded by a roundish envelope. $e_4$ is $> 0$ at the transition.

1177 (= RB 045): The evidence for a disk is marginal.

1181 (= RB 043): The isophotes in the envelope are strongly twisted compared to the main body of this small boE object, hence its classification as pec. This is reminiscent of NGC 4478 in the Virgo cluster. Our CCD TBL image does not confirm this twist.

1233 (= NGC 4874): It is the western component of the characteristic pair of cD galaxies in the Coma cluster (with NGC 4889). It displays an extended envelope, well above the $r^{1/4}$ law, as extrapolated from the inner regions.

1258 (= RB 026): Difficult analysis, due to the halo of a very bright star.

1266 (= RB 234): The evidence for a disk is marginal.

1277 (= RB 022): Difficult analysis, due to the halo of a very bright star.

1373 (= RB 268): This roundish galaxy has such a low effective SuBr that it is probably a disk dominated object.

1378 (= NGC 4864): A rather bright star 7 arcsec away from the galaxian center makes the isophotal analysis rather uncertain.

1414 (= IC 3955): See description in Sect. 5.1.

1458 (= IC 3949): Edge-on galaxy with an important dust pattern.

1489 (= RB 252): A strongly inclined SA0 with evidence for an $f_4$-asymmetry.

1620 and 1623 (= NGC 4851): Pair of mildly overlapping galaxies.

1724: This is not one galaxy, but two overlapping galaxies with their nuclei about 3 arcsec apart.

1739: Doubtful case! $e_4$ is clearly $> 0$ only for a $< 3$", i.e. in a range where it is uncertain.

1758: This object becomes roundish outwards. The envelope might be a disk with an unresolved spiral pattern or a spheroidal halo. The latter is less likely because of the large and consistent twist of the outer isophotes.

1844: Important ring-lens.

1687: Important dust pattern.

1778: An edge-on S0 with a clear $f_4$-asymmetry.

2034: Important dust pattern. This object might be an Sa.

*5.4 KP2225 sample*

D16: This galaxy is asymmetric and of low SuBr with evidence for dust.

FW10: A disk is detected at large radii.

NGC 4692: This galaxy has been classified as unE from our CCD image, at variance with the present determination. The twist of the PA is uncertain for this roundish galaxy.

FW2: The twist of the PA is uncertain.

D3: An edge-on galaxy with a clear $f_4$-asymmetry.

anon04: Galaxy interacting with supp08.

D11: An edge-on galaxy with a clear $f_4$-asymmetry.

D10: This galaxy has a ring similar to that of NGC 4215.

FW9: This unE shows an outer disk.

D12: An edge-on SA0 with a clear $f_4$-asymmetry.

supp01: The PA twist is uncertain.

supp06: Dust detected.

D14: A low SuBr dusty galaxy.

supp08: It is interacting with anon04.

D6: A spiral, judging from the asymmetry.

supp09: Asymmetric and dusty galaxy.

supp12: This galaxy is not FW3.

## 6. Internal consistency and comparison with other studies

*6.1 TBL vs CFHT*

There are 13 galaxies in common between the TBL and CFHT subsamples. Ten of them are assigned the same morphological type in the respective tables (Tables 2a and 2b). The better data from CFHT allow us to classify the unE GMP1878 as diE, and the unE/SA0 GMP1823 as SA0. Because of the small field of view, the photometric disk of the bright galaxy GMP1878 was missed on the CFHT image, and this SA0 galaxy was classified diE. The discrepancies are thus well understood, and we consider that there is a perfect agreement between the classifications derived from the TBL and CFHT images.

*6.2 CCD vs plates*

Differences in classification between CCD and plates essentially result from differences in seeing, since most CCD images have better resolution than the plates, and from density to intensity conversion errors, which may affect the photometric profiles. There are 35 galaxies in common between the CCD and KP1608 samples. A good agreement is found for 33 galaxies; 21 of them have the same morphological label in both samples, and the 12 others belong to a different subclass of the *same* class in both samples (for example, some SAB0's in one sample are classified SB0's in the other). The external isophotes of GMP 1210 are irregular; this is clearly seen on the CCD image, but could be attributed to noise on the plate image, hence the slightly different classification. The other discrepant galaxy, GMP 1300, was classified SAa from the KP1608 plate, but no evidence for a spiral pattern was found on the CCD data, which lead to SA0 type. This fully justifies using photographic plates for classifying galaxies at the distance of Coma.

Effective radii for the galaxies in common may differ in some cases; $r_e$(plate) tend to be larger than $r_e$(CCD) for smaller objects, presumably because of seeing and of density-to-intensity conversion errors. Thus the ellipticities and $e_4$, measured at the effective radius, may differ. But this has no consequences on the morphological classification, which is the main objective of this paper.

*6.3 Comparison with other studies : asymptotic magnitudes*

We have compared our asymptotic magnitudes with the isophotal magnitudes given by GMP for the TBL subsample and for the KP1608 sample. GMP did not measure the FW sample, and, as mentioned in Sect. 4.1, the magnitudes for the CFH subsample are not very reliable. GMP actually give blue magnitudes at the isophote level 26.5, and a color (B − r) at a given radius, from which we derived $r_{GMP}$.

The results are the following. Our asymptotic magnitudes in R are on the average one tenth of a magnitude fainter than GMP (the difference $r_{GMP}$ − $R_{us}$ is − 0.09 for the TBL subsample and − 0.10 for the KP1608 sample). The zero-point difference probably results from the difference between GMP's filter F (which they call r) and the standard Cousin's R. GMP give a scatter of 0.15 in the isophotal magnitude, from which we derive that our uncertainty is also about 0.15 (since the scatter between their magnitudes and ours is 0.21). This result is satisfactory, since our goal is not accurate photometry, but classification.

*6.4 Comparison with other studies : ellipticity and $e_4$*

Figure 4 shows the comparison between our values for the ellipticity and the $e_4$ parameter with those of Jørgensen & Franx (1994, hereafter JF) and Saglia et al. (1993, hereafter SBD) for galaxies in common. The two parameters $\epsilon$ and $e_4$ refer to slighty different quantities in the three papers. In JF, the ellipticity $\epsilon = (1 - b/a)$ is determined at the effective radius. SBD do not specify where the ellipticity is measured, but they probably measured it at the effective radius, like JF, because they agree with JF for galaxies in common, and disagree with us for galaxies with large ellipticity variations. We determine the axis ratio $b/a$ at its extremum value, if well defined, and at the effective radius otherwise. Our axis ratio has been converted to an ellipticity for the purpose of comparison.

In JF, the $e_4$ parameter (called $c_4$) is the extremum of the radial profile of $e_4$, if it is present; otherwise $e_4$ is computed at the effective radius, as in our paper. In SBD, the $e_4$ parameter (called $a4/a$) is the value at the extremum, if present; otherwise it is 0.

Moreover, the ellipticity and $e_4$ profiles cover different galaxy radial ranges in the three studies, because of different seeing conditions and S/N ratios; our CCD images and SBD's generally have longer exposure times and better seeing than those of JF. Finally, the three teams use slightly different ways to compute these two parameters. In particular, Michard & Marchal (1994a) have already pointed out that their technique of isophotal analysis, used in the present work, gives larger absolute values for $e_4$ than the method of Bender and others.

The top panels of Fig. 4 show the comparison of the ellipticities measured by the three teams. We measure systematically larger values for the ellipticity, as expected from our definition of $\epsilon$.

The three most discrepant galaxies are NGC 4908, IC 4051 and RB 099 (triangles above the diagonal on Figs. 4a and b). The ellipticity profiles of these galaxies monotonously increase outward, and the effective radii given in Tables 2a and 2b are measured at the effective radius. Our values for the effective radii do not agree with those of JF, thus explaining the discrepancy. At the effective radii adopted by JF, we do measure the same ellipticities as them.

Two other galaxies (square above the diagonal on Figs. 4 a and b) have slightly larger values of ellipticity in the literature than in our Table 2b. These are two large galaxies which we imaged with a small-field CCD at CFH; our value should thus be considered an upper limit.

19 galaxies of our TBL subsample were also measured by Jørgensen et al. (1992), who give radial profiles, allowing a detailed comparison. We find $\epsilon_{us} = 1.034 \times \epsilon_{JF} - 0.004$, with a correlation coefficient of 0.97. Similar results are found when comparing our ellipticities at $\mu_R = 21.60$ with those of JF at $\mu_r = 21.85$ (r Gunn), which corresponds roughly to the same isophote level, and our elllipticities at $\mu_R = 24.$ with those of GMP at $\mu_R = 24.$

The lower panels of Fig. 4 show the comparison of the $e_4$ parameters measured by the three teams. A correlation between the values is certainly present in both comparisons. The agreement is better with JF than with SBD. The scatter is large; it can probably be accounted for by the different methods for measuring $e_4$, and by galaxies whose $e_4$ is not well defined, such as SB0's and roundish E's.

*6.5 Comparison with other studies : morphological classification*

Original morphological classifications of large samples of Coma cluster galaxies have been published by Rood & Baum (1967), and by Dressler (1980). In Dressler's list there are 140 galaxies of early-type (the latest being S0/a) which we also classified, as they fall either in our CCD or KP1608 sample. These galaxies were used for a systematic comparison. Since the criteria of our "quantitative morphology", summarized in Sect.

3, are not quite the same as those of classical morphology, we may expect some systematic differences. They indeed occur:

– of the 54 objects seen as ellipticals by Dressler, that is D + E + E/S0 together, only 36 are still E in our classification, 18 have been shifted to the S0 classes, and no less than 11 to the SAB0 and SB0 cells.

– of the 83 lenticulars of Dressler, that is his S0/E, S0 and SB0, we put 7 in the E class, and 8 in the S0/a transition class. This might look like a good agreement, except that we detect bars in 19 of Dressler's unbarred S0's. There seems to be a strong deficit of SB0's in Dressler's morphology, also in comparison with Rood and Baum, as noted also by SBD. Inspection of Dressler's catalogue of morphological types of galaxies in clusters shows that Dressler's S0 class contains all lenticular that have no bar, as well as lenticulars for which nothing can be said about the presence of a bar at the resolutions of his data. This explains the deficit of SB0's.

– to be complete, there are only 3 transition cases S0/a in Dressler's data: we spread these 3 cases into 3 adjoining cells, including one Sa.

A similar comparison based on the smaller, but complete, KP1608 sample was also performed (Michard, 1995). The results are in agreement: 1/3 of Dressler's ellipticals show photometric evidence for a disk, and nearly 25% of his E's and non barred S0's show evidence for a bar from the measured isophotal twists. These differences are due to the different techniques for ascertaining the presence of a disk or bar. It may be added that more than 1/3 of Dressler's ellipticals are classified as diE's from the $e_4$ criterion alone, so that the disk remains undetected in less than 1/3 of "classical" E galaxies, that is in 11% of the total early-type population of the cluster.

Finally, the comparison of morphological types in a magnitude-complete sample of about 200 galaxies in Coma (Andreon et al., in preparation) with the present and with published samples gives a good agreement for almost all galaxies, once the differences due to the classification methods are taken into account.

**Acknowledgements.** We thank Guy Monnet for participating in the CFHT run, Jacques Marchal for taking the CCD images of several runs, Hervé Wozniak for taking some of the CCD images in the last TBL run, and the staffs of the TBL and CFH telescopes for assistance during the observations. A warm thank goes to Dr. Schoening who gave us access to the Kitt Peak collection of Coma plates. We thank Jean Guibert and his staff for digitizing the KPNO plates, and Albert Bijaoui, Jacques Marchal, and Marc Giudicelli for contributing to the reduction software. SA thanks prof. G. Boella, Director of the Istituto di Fisica Cosmica del CNR, were part of this work was done, and acknowledges fellowships from the University of Trieste and from the French Ministry of Foreign Affairs. All the tables (including those for the photometric et geometrical profiles of the individual galaxies as a function of radius) will be made available in electronic form at CDS in Strasbourg.

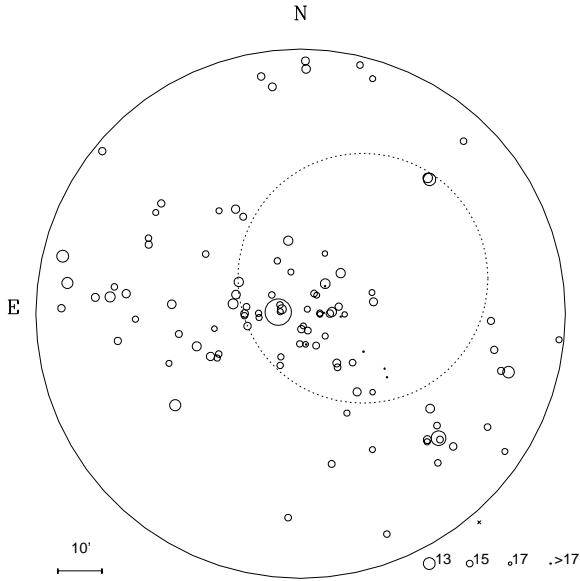

**Fig. 1.** The galaxies of the CCD sample. The size of the open circles is proportional to the magnitude of the galaxy. The radius of the field is one degree. The dashed circle shows the extent of the KP1608 sample.

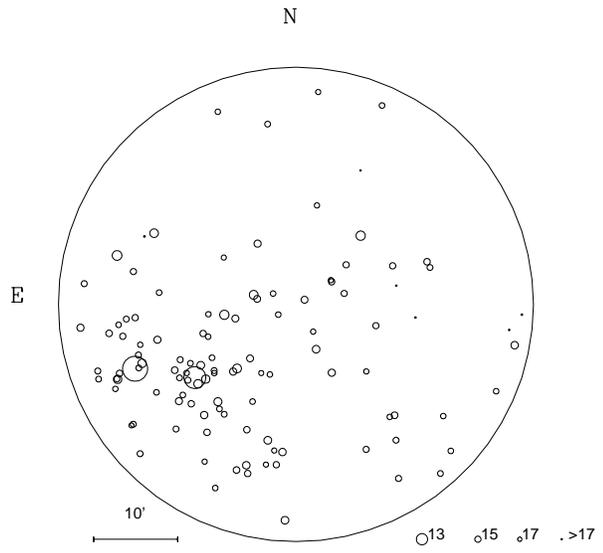

**Fig. 2.** The galaxies of plate KP1608 (the KP1608 sample). The size of the open circles is proportional to the magnitude of the galaxy. The diameter of the field is 50 arcmin. The center of the field is located 7.5 arcmin North and 19.4 arcmin West of that of the cluster.

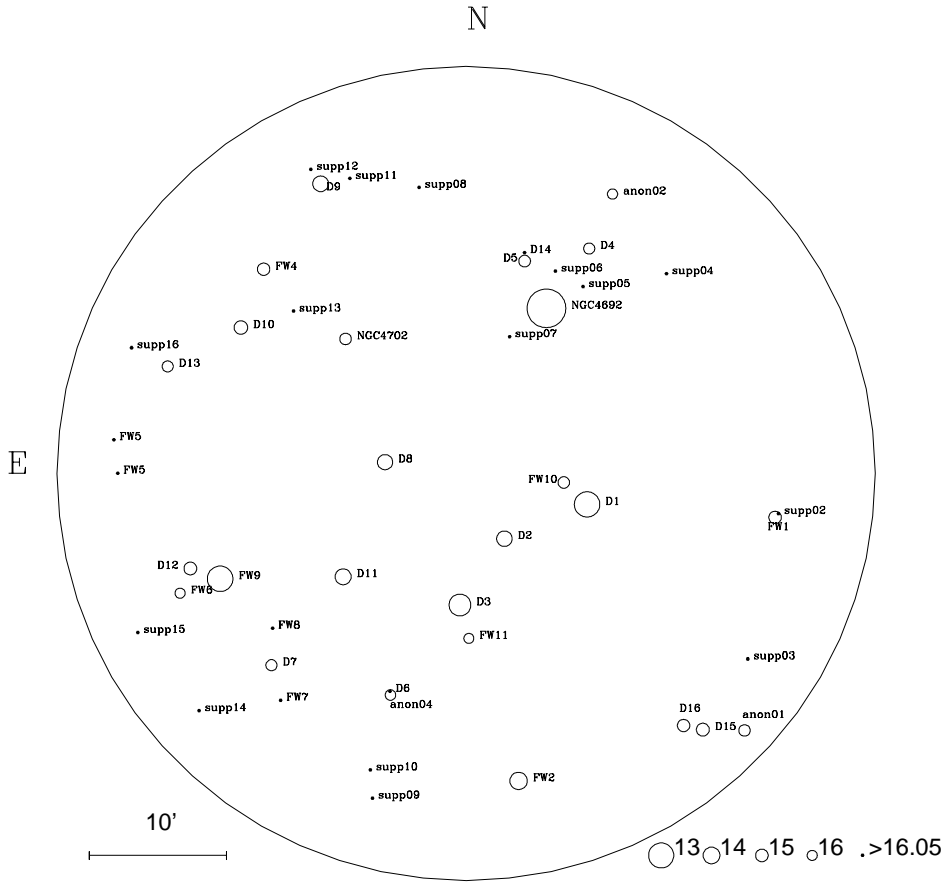

**Fig. 3.** The galaxies of plate KP2225 (FW sample). The size of the open circles is proportional to the magnitude of the galaxy. The diameter of the field is 50 arcmin. It is located 2.6 degrees West of the center of Coma.

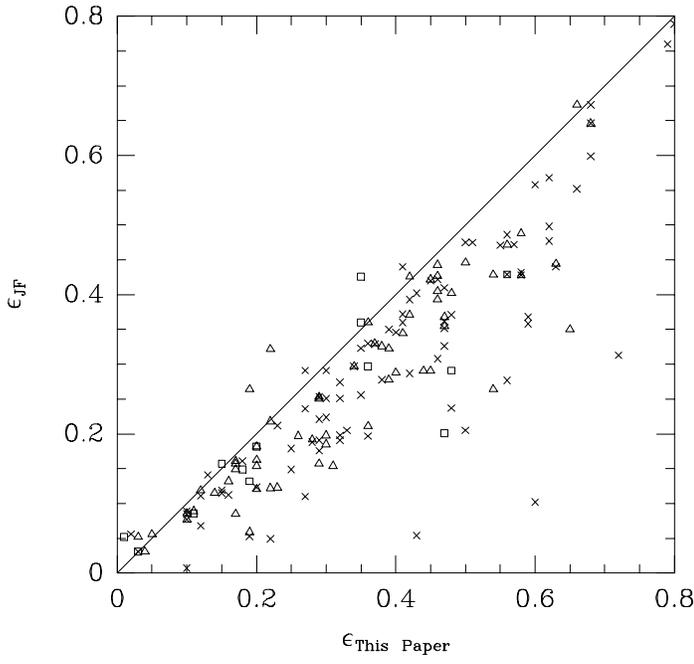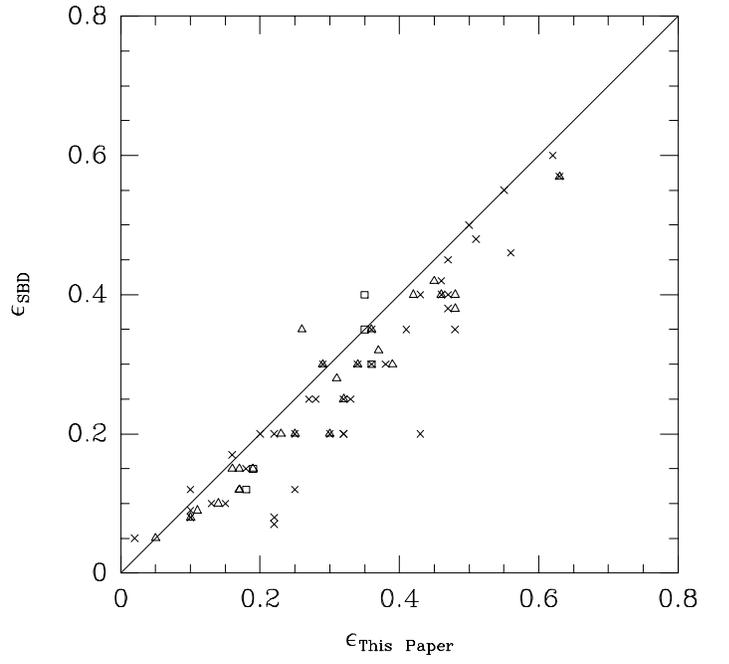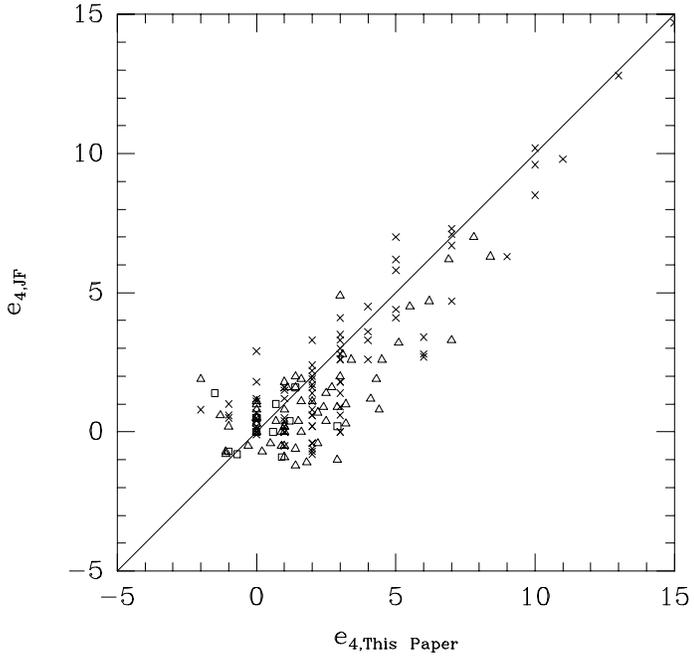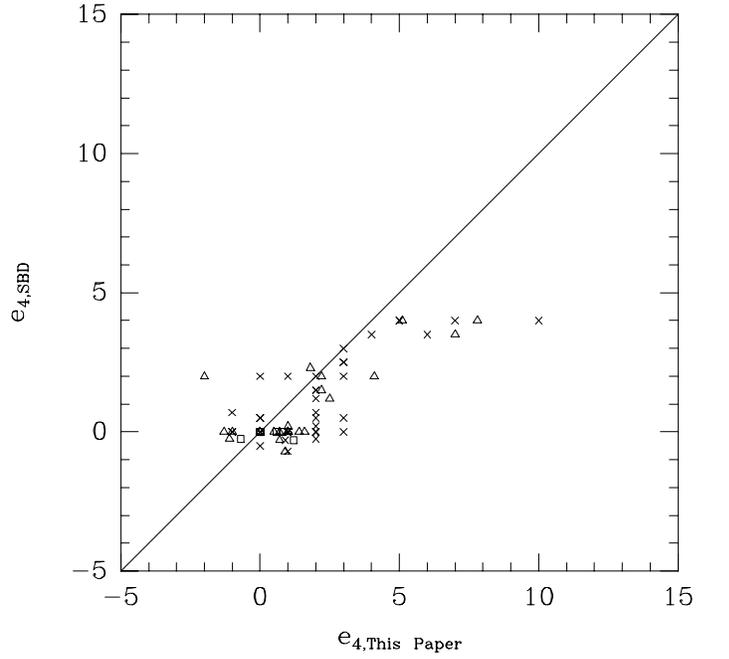

**Figure 4.** Comparison of the ellipticities $\epsilon$ obtained in this paper (upper panels) with those of JF (left) and those of SBD, and of the $e_4$ obtained in this paper (bottom panels) with those of JF (left) and of SBD (right). Triangles refer to the TBL subsample, squares to the CFH subsample, and crosses to the KP1608 sample.

Table 2a. Photometric parameters and classification of the CCD TBL subsample

| (1) | (2) | (3) | (4) | (5) | (6) | (7) | (8) | (9) | (10) | (11) | (12) | (13) | (14) | (15) | (16) | (17) | (18) | (19) | (20) | (21) | (22) |
|---|---|---|---|---|---|---|---|---|---|---|---|---|---|---|---|---|---|---|---|---|---|
| 0231 | 0739 |     |         | -3253 |   75 | 13.73 | 0.74 | 0.75 | 10.53 | no | 0.97 | -0.3 | re | 0.87 | 45 | -no  | -no-  | -no-  | -?-  | unE      |   |
| 0238 | 0756 |     | NGC4944 | -3236 |  783 | 12.33 | 0.89 | 1.09 |  9.88 | cl | 0.29 |  1.2 | ex | 0.45 | 03 | -no  | emDi  | -no-  | thD  | SA0      | * |
| 0253 | 0798 |     | NGC4943 | -3172 |  418 | 13.93 | 0.59 | 0.65 |  9.97 | st | 0.55 |  1.0 | co | 0.76 | 15 | bar? | emDi  | -no-  | spH  | SAB0     |   |
| 0334 | 1046 |     | NGC4934 | -2792 |  222 | 13.51 | 0.66 | 0.95 |  9.90 | st | 0.21 |  4.5 | re | 0.25 | 00 | -no  | miDi  | -no-  | thD  | SA0      | * |
| 0352 | 1111 |     |         | -2697 | 2212 | 13.53 | 0.49 | 0.61 |  9.07 | st | 0.30 | 10.8 | ex | 0.40 | 00 | -no  | emDi  | -no-  | thD  | SA0      |   |
| 0375 | 1176 | 164 | NGC4931 | -2591 |  229 | 12.69 | 0.67 | 0.93 |  9.15 | cl | 0.37 |  9.5 | co | 0.45 | 03 | -no  | miDi  | -no-  | thD  | SA0      |   |
| 0390 | 1223 | 165 |         | -2532 |  365 | 14.27 | 0.70 | 0.85 | 10.98 | cl | 0.47 |  2.1 | ex | 0.50 | 00 | -no  | emDi  | -no-  | thD  | SA0      |   |
| 0421 | 1322 | 166 | NGC4929 | -2372 |  273 | 13.22 | 1.02 | 1.07 | 11.41 | no | 0.84 |  0.4 | re | 0.76 | 19 | -no  | -no-  | -no-  | pec  | unE      | * |
| 0453 | 1397 | 113 |         | -2246 |  -74 | 14.53 | 0.81 | 0.94 | 11.66 | st | 0.45 |  2.0 | ex | 0.48 | 35 | bar  | exDi  | -no-  | exD  | SB0      |   |
| 0484 | 1503 | 214 |         | -2069 | 1029 | 14.15 | 0.61 | 0.73 | 10.33 | st | 0.52 |  2.1 | ex | 0.64 | 02 | -no  | miDi  | -no-  | thD  | SA0      |   |
| 0485 | 1504 | 213 |         | -2066 |  940 | 13.79 | 0.48 | 0.62 |  9.27 | st | 0.36 |  5.5 | co | 0.48 | 05 | -no  | miDi  | spiP  | thD  | Sa       | * |
| 0504 | 1566 | 227 |         | -1970 | 1378 | 14.67 | 0.90 | 1.01 | 12.27 | ft | 0.66 | -3.4 | ex | 0.58 | 28 | bar  | -?Di  | spiP? | exD  | SAB0/a   | * |
| 0517 | 1614 | 228 |         | -1895 | 1501 | 13.75 | 0.69 | 0.80 | 10.27 | cl | 0.54 |  6.1 | ex | 0.75 | 29 | bar  | -?Di  | -no-  | -?-  | SB0      |   |
| 0539 | 1681 | 061 |         | -1790 | -675 | 14.28 | 0.34 | 0.56 |  9.07 | st | 0.29 | 11.5 | ex | 0.35 | 04 | -no  | miDi  | -no-  | thD  | SA0      | * |
| 0552 | 1715 | 141 | NGC4927 | -1752 |  129 | 12.85 | 0.91 | 0.98 | 10.48 | no | 0.72 |  1.5 | ex | 0.65 | 00 | -no  | -?Di  | -no-  | -?-  | diE      |   |
| 0565 | 1750 | 048 | NGC4926 | -1705 |-1245 | 12.21 | 0.95 | 0.99 | 10.07 | no | 0.84 |  1.0 | co | 0.85 | 11 | -no  | emDi  | -no-  | -?-  | diE      |   |
| 0591 | 1807 | 096 |         | -1656 | -276 | 13.92 | 0.55 | 0.58 |  9.76 | no | 0.88 |  1.7 | ex | 0.94 | 25 | bar? | -?Di  | -no-  | -?-  | diE/SAB0 | * |
| 0661 | 2000 | 078 | NGC4923 | -1413 | -443 | 12.97 | 0.80 | 0.83 | 10.05 | no | 0.80 |  0.0 | co | 0.80 | 50 | -no  | -no-  | -no-  | -?-  | unE      | * |
| 0694 | 2091 | 204 |         | -1291 |  812 | 14.24 | 0.70 | 0.77 | 10.82 | no | 0.72 |  0.0 | re | 0.63 | 05 | -no  | -?Di  | -no-  | pec  | Epec     | * |
| 0718 | 2157 | 079 | NGC4919 | -1225 | -581 | 13.22 | 0.78 | 0.88 | 10.20 | st | 0.53 |  3.2 | ex | 0.52 | 04 | -no  | exDi  | -no-  | exD  | SA0      | * |
| 0750 | 2237 | 080 |         | -1134 | -603 | 14.44 | 0.59 | 0.66 | 10.50 | cl | 0.64 |  2.4 | ex | 0.94 | 25 | -no  | emDi  | -no-  | spH  | SA0      |   |
| 0754 | 2252 | 081 |         | -1114 | -548 | 14.42 | 0.64 | 0.69 | 10.74 | no | 0.70 |  2.5 | ex | 0.70 | 18 | -no  | -?Di  | -no-  | -?-  | diE      |   |
| 0757 | 2259 | 229 |         | -1109 | 1401 | 14.64 | 0.59 | 0.63 | 10.70 | ft | 0.78 |  1.4 | co | 0.83 | 95 | bar  | -?Di  | -no-  | -?-  | SB0      |   |
| 0812 | 2390 | 143 | IC4051  |  -918 |  134 | 12.77 | 1.03 | 1.08 | 11.06 | no | 0.81 | -0.3 | ex | 0.66 | 15 | -no  | -no-  | -no-  | -?-  | unE      |   |
| 0825 | 2413 | 230 |         |  -884 | 1423 | 13.17 | 0.72 | 0.77 |  9.88 | st | 0.54 |  1.6 | ex | 0.61 | 50 | bar  | -no-  | emDi  | spH  | SB0      |   |
| 0828 | 2417 | 167 | NGC4908 |  -879 |  260 | 12.98 | 0.73 | 0.79 |  9.73 | no | 0.78 |  1.0 | co | 0.60 | 10 | -no  | -?Di  | -no-  | -?-  | diE      |   |
| 0836 | 2440 | 168 | IC4045  |  -841 |  433 | 13.52 | 0.50 | 0.60 |  9.13 | ft | 0.63 |  2.5 | ex | 0.77 | 04 | -no  | emDi  | -no-  | spH  | diE/SA0  |   |
| 0857 | 2495 | 231 |         |  -781 | 1320 | 13.81 | 0.50 | 0.59 |  9.42 | cl | 0.54 |  4.3 | ex | 0.61 | 13 | bar? | emDi  | -no-  | thD  | SAB0     | * |
| 0863 | 2510 | 116 |         |  -764 |  -27 | 14.35 | 0.56 | 0.65 | 10.27 | cl | 0.69 |  5.1 | co | 0.78 | 53 | bar  | -?Di  | -no-  | -?-  | SB0      |   |
| 0867 | 2516 | 144 | IC4042  |  -762 |    3 | 13.54 | 0.62 | 0.66 |  9.75 | cl | 0.81 |  1.8 | ex | 0.84 | 70 | bar? | -?Di  | spiP? | -?-  | SA0/a    | * |
| 0874 | 2535 | 145 | IC4041  |  -736 |   94 | 14.09 | 0.68 | 0.79 | 10.60 | cl | 0.54 |  2.2 | co | 0.75 | 05 | -no- | emDi  | -no-  | spH  | SA0      |   |
| 0876 | 2541 | 118 | NGC4906 |  -722 | -168 | 13.57 | 0.72 | 0.76 | 10.27 | no | 0.80 |  0.9 | ex | 0.96 | 15 | -no  | -?Di  | -no-  | -?-  | unE      | * |
| 0917 | 2651 | 147 | RB100   |  -571 |    6 | 14.32 | 0.66 | 0.80 | 10.73 | st | 0.37 |  7.0 | ex | 0.60 | 00 | -no  | emDi  | -no-  | spH  | SA0      |   |
| 0918 | 2654 | 119 | RB099   |  -565 |  -52 | 14.40 | 0.39 | 0.43 |  9.46 | –  | 0.74 |  4.1 | ex | 0.61 | 42 | bar  | -?Di  | -no-  | -?-  | SB0      |   |
| 0930 | 2682 |     |         |  -536 | 3228 | 13.83 | 0.73 | 0.81 | 10.60 | no | 0.67 | -0.4 | re | 0.73 | 04 | -no  | -no-  | -no-  | -?-  | unE      |   |
| 0999 | 2839 | 172 | IC4021  |  -392 |  255 | 14.27 | 0.43 | 0.45 |  9.50 | ft | 0.83 |  1.6 | ex | 0.86 | 00 | -no  | -?Di  | -no-  | -?-  | SA0      |   |
| 1004 | 2847 |     |         |  -384 | 3087 | 13.84 | 0.88 | 1.00 | 11.04 | cl | 0.48 |  1.1 | co | 0.50 | 04 | -no  | exDi  | -no-  | exD  | SA0      |   |
| 1035 | 2912 | 207 | RB167   |  -316 |  719 | 14.40 | 0.58 | 0.66 | 10.46 | ft | 0.62 | -2.0 | ex | 0.66 | 05 | bar? | -?Di  | -no-  | -?-  | SAB0     | * |
| 1039 | 2921 | 148 | NGC4889 |  -304 |   22 | 11.10 | 1.26 | 1.35 | 10.48 | no | 0.64 | -1.1 | ex | 0.62 | 00 | -no  | -no-  | -no-  | pec  | boE      | * |
| 1046 | 2940 | 150 | IC4011  |  -280 |  120 | 14.57 | 0.54 | 0.57 | 10.37 | ft | 0.89 | -1.3 | ex | 0.90 | 10 | -no  | -no-  | -no-  | -?-  | boE      |   |
| 1049 | 2945 | 065 |         |  -278 | -703 | 14.47 | 0.64 | 0.74 | 10.78 | cl | 0.42 |  4.4 | ex | 0.49 | 02 | -no  | emDi  | -no-  | thD  | SA0      |   |
| 1050 | 2946 | 149 | RB077   |  -277 |   28 | 14.98 | 0.44 | 0.61 | 10.26 | st | 0.34 |  6.2 | ex | 0.57 | 04 | -no  | emDi  | -no-  | spH  | SA0      | * |
| 1057 | 2956 | 084 |         |  -268 | -587 | 13.87 | 0.51 | 0.66 |  9.51 | st | 0.32 |  6.9 | ex | 0.42 | 00 | -no  | emDi  | -no-  | thD  | SA0      |   |
| 1065 | 2975 | 151 | NGC4886 |  -254 |   61 | 13.26 | 0.83 | 0.84 | 10.49 | no | 0.96 |  0.0 | re | 0.95 | 25 | -no  | -no-  | -no-  | -?-  | unE      | * |
| 1102 | 3053 |     |         |  -170 |-2775 | 14.07 | 0.75 | 0.82 | 10.91 | no | 0.75 |  0.0 | re | 0.77 | 05 | -no  | -no-  | -no-  | -?-  | unE      |   |

**Table 2a.** Continued

| (1) | (2) | (3) | (4) | (5) | (6) | (7) | (8) | (9) | (10) | (11) | (12) | (13) | (14) | (15) | (16) | (17) | (18) | (19) | (20) | (21) | (22) |
|---|---|---|---|---|---|---|---|---|---|---|---|---|---|---|---|---|---|---|---|---|---|
| 1104 | 3055 | 217 | NGC4881 | -168 | 994 | 12.88 | 0.91 | 0.93 | 10.54 | no | 0.95 | -1.0 | ex | 0.95 | 40 | -no | -no- | -no- | -?- | boE | |
| 1115 | 3084 | 193 | RB155 | -132 | 568 | 14.32 | 0.43 | 0.47 | 9.61 | cl | 0.77 | 2.2 | ex | 0.98 | 35 | bar | -?Di | -no- | -?- | SB0 | * |
| 1164 | 3178 | 101 | RB049 | -11 | -410 | 14.30 | 0.47 | 0.54 | 9.75 | cl | 0.35 | 8.4 | ex | 0.35 | 00 | -no | exDi | -no- | exD | SA0 | |
| 1172 | 3201 | 124 | NGC4876 | 10 | -210 | 13.63 | 0.55 | 0.62 | 9.49 | ft | 0.66 | 1.0 | ex | 0.74 | 06 | -no | emDi | -no- | -?- | diE | |
| 1181 | 3222 | 125 | RB043 | 38 | -166 | 14.84 | 0.23 | 0.26 | 9.09 | no | 0.88 | -1.0 | ex | 0.89 | 05 | -no | -no- | -no- | -?- | boE | |
| 1197 | 3259 | | | 66 | 3440 | 13.45 | 0.72 | 0.88 | 10.15 | st | 0.47 | 9.7 | ex | 0.62 | 02 | -no | emDi | -no- | spH | SA0 | |
| 1198 | 3262 | 102 | RB041 | 68 | -417 | 15.28 | 0.37 | 0.46 | 10.21 | no | 0.59 | 2.7 | ex | 0.68 | 00 | -no | emDi | -no- | -?- | diE | * |
| 1204 | 3273 | | | 75 | 3330 | 12.91 | 0.74 | 0.82 | 9.72 | st | 0.27 | 5.3 | ex | 0.36 | 12 | bar? | emDi | -no- | thD | SAB0/a | * |
| 1210 | 3291 | 154 | RB038 | 92 | 61 | 14.90 | 0.71 | 0.79 | 11.53 | cl | 0.61 | 1.4 | ex | 0.67 | 10 | -no | -?Di | spiP? | -?- | SA0/a | |
| 1214 | 3296 | 104 | NGC4875 | 98 | -229 | 14.03 | 0.41 | 0.45 | 9.15 | cl | 0.70 | 1.1 | ex | 0.81 | 12 | bar? | emDi | -no- | -?- | SAB0 | * |
| 1258 | 3390 | 176 | RB026 | 183 | 275 | 14.28 | 0.42 | 0.52 | 9.49 | cl | 0.44 | 5.5 | ex | 0.70 | 00 | -no | emDi | -no- | spH | SA0 | * |
| 1274 | 3423 | 088 | IC3976 | 211 | -434 | 13.91 | 0.36 | 0.43 | 8.79 | st | 0.42 | 4.5 | ex | 0.60 | 00 | -no | emDi | -no- | spH | SA0 | |
| 1277 | 3433 | 177 | RB022 | 219 | 252 | 14.82 | 0.57 | 0.68 | 10.74 | no | 0.54 | 1.0 | ex | 0.67 | 20 | -no | emDi | -no- | -?- | diE | * |
| 1300 | 3484 | 157 | RB014 | 262 | 9 | 14.66 | 0.45 | 0.51 | 10.03 | cl | 0.56 | 2.9 | ex | 0.59 | 10 | -no | -?Di | -no- | -?- | SA0 | * |
| 1302 | 3487 | 132 | RB013 | 263 | -9 | 14.88 | 0.54 | 0.59 | 10.69 | cl | 0.58 | 3.4 | ex | 0.71 | 17 | -no | emDi | -no- | spH | SA0 | |
| 1327 | 3553 | 208 | RB136 | 331 | 819 | 14.15 | 0.56 | 0.65 | 10.02 | cl | 0.60 | 2.0 | ex | 0.63 | 05 | -no | exDi | -no- | exD | SA0 | |
| 1329 | 3557 | 107 | RB006 | 334 | -304 | 14.72 | 0.49 | 0.63 | 10.29 | ft | 0.52 | 0.9 | ex | 0.83 | 12 | -no | emDi | -no- | spH | diE/SA0 | |
| 1331 | 3561 | 179 | NGC4865 | 335 | 409 | 13.49 | 0.67 | 0.78 | 9.93 | ft | 0.46 | 1.4 | ex | 0.51 | 00 | -no | emDi | -no- | -?- | diE | |
| 1365 | 3639 | 133 | NGC4867 | 398 | 2 | 13.77 | 0.53 | 0.59 | 9.52 | ft | 0.71 | 1.0 | ex | 0.83 | 00 | -no | emDi | -no- | spH? | diE | |
| 1375 | 3661 | 013 | | 423 | -2044 | 13.91 | 0.44 | 0.51 | 9.23 | cl | 0.60 | 4.2 | co | 0.77 | 08 | -no | miDi | -no- | spH | SA0 | |
| 1378 | 3664 | 159 | NGC4864 | 426 | 24 | 12.77 | 0.83 | 0.87 | 10.04 | no | 0.83 | 0.0 | co | 0.92 | 11 | -no | -no- | -no- | -?- | unE | * |
| 1401 | 3730 | 069 | IC3959 | 492 | -670 | 13.60 | 0.62 | 0.65 | 9.65 | no | 0.86 | 0.0 | co | 0.96 | 00 | -no | -no- | -no- | -?- | unE | |
| 1406 | 3739 | 070 | IC3957 | 502 | -729 | 14.14 | 0.53 | 0.56 | 9.90 | no | 0.90 | 0.5 | re | 0.92 | 15 | -no | -no- | -no- | -?- | unE | * |
| 1414 | 3761 | 160 | IC3955 | 519 | 96 | 13.84 | 0.73 | 0.77 | 10.58 | st | 0.55 | 7.8 | ex | 0.80 | 32 | bar | -?Di | spiP | -?- | SB0/a | * |
| 1425 | 3792 | 194 | NGC4860 | 545 | 552 | 12.78 | 0.83 | 0.87 | 10.03 | no | 0.83 | 1.0 | co | 0.93 | 06 | -no | -?Di | -no- | -?- | diE | |
| 1444 | 3851 | 135 | RB260 | 599 | -9 | 15.21 | 0.44 | 0.50 | 10.50 | no | 0.75 | 0.7 | ex | 0.80 | 08 | -no | -?Di | -no- | -?- | diE | |
| 1453 | 3879 | 042 | | 630 | -1352 | 14.41 | 0.32 | 0.38 | 9.13 | cl | 0.53 | 3.0 | re | 0.70 | 10 | bar? | miDi | -no- | spH | SAB0 | |
| 1485 | 3958 | 072 | IC3947 | 707 | -666 | 14.38 | 0.47 | 0.56 | 9.81 | ft | 0.61 | 3.1 | ex | 0.92 | 03 | -no | emDi | -no- | spH | SA0 | |
| 1507 | 4017 | 058 | NGC4854 | 769 | -1064 | 13.21 | 1.15 | 1.21 | 12.05 | ft | 0.71 | 1.5 | ex | 0.74 | 32 | bar? | -?Di | -no- | -?- | SAB0 | * |
| 1524 | 4044 | | | 808 | 3384 | 14.44 | 0.77 | 0.82 | 11.37 | cl | 0.80 | 1.7 | ex | 0.84 | 90 | bar | -?Di | -no- | -?- | SBa | * |
| 1581 | 4200 | 182 | RB243 | 971 | 288 | 15.18 | 0.31 | 0.35 | 9.82 | ft | 0.83 | 1.1 | ex | 0.97 | 16 | bar? | emDi | -no- | -?- | SAB0 | |
| 1584 | 4206 | 020 | | 977 | -1849 | 14.15 | 0.43 | 0.53 | 9.40 | cl | 0.50 | 1.6 | ex | 0.71 | 06 | -no | emDi | -no- | spH | SA0 | |
| 1585 | 4209 | 059 | RB188 | 979 | -1067 | 15.17 | 0.21 | 0.28 | 9.32 | no | 0.71 | 3.2 | ex | 0.97 | 05 | -no | -?Di | -no- | -?- | diE | |
| 1588 | 4213 | | | 980 | 3199 | 14.77 | 0.66 | 0.79 | 11.15 | cl | 0.50 | 7.3 | co | 0.99 | 70 | -no | emDi | spiP? | pec | SA0/a | |
| 1594 | 4230 | 161 | RB241 | 992 | 162 | 13.28 | 0.90 | 0.93 | 10.89 | no | 0.88 | 0.0 | co | 0.96 | 04 | -no | -no- | -no- | -?- | unE | |
| 1646 | 4362 | | | 1173 | -2999 | 14.10 | 0.90 | 0.97 | 11.71 | no | 0.74 | 3.1 | ex | 0.73 | 17 | bar? | -?Di | -no- | -?- | Irr | * |
| 1821 | 4792 | 023 | NGC4842b | 1721 | -1743 | 14.47 | 0.37 | 0.44 | 9.44 | st | 0.55 | 3.0 | ex | 0.78 | 05 | -no | miDi | -no- | spH | SA0 | |
| 1823 | 4794 | 030 | NGC4842a | 1724 | -1712 | 13.43 | 0.52 | 0.55 | 9.12 | ft | 0.90 | 0.0 | co | 0.98 | 09 | -no | -no- | -no- | -?- | unE/SA0 | * |
| 1828 | 4806 | 239 | NGC4841b | 1729 | 1848 | 12.91 | 0.91 | 0.91 | 10.56 | no | 0.97 | 0.0 | re | 0.93 | – | -no | -no- | -no- | -?- | unE | * |
| 1831 | 4822 | 240 | NGC4841a | 1754 | 1828 | 11.91 | 1.19 | 1.23 | 10.97 | no | 0.83 | -1.1 | co | 0.85 | 36 | -no | -no- | -no- | -?- | boE | * |
| 1834 | 4829 | 046 | NGC4840 | 1762 | -1291 | 13.02 | 0.72 | 0.77 | 9.71 | ft | 0.80 | 1.0 | ex | 0.88 | 15 | bar? | -?Di | -no- | -?- | SAB0 | |
| 1869 | 4907 | 033 | | 1855 | -1522 | 14.08 | 0.35 | 0.40 | 8.92 | cl | 0.60 | 2.0 | ex | 0.68 | 01 | -no | emDi | -no- | spH | SA0 | |
| 1874 | 4918 | 015 | | 1867 | -2030 | 14.93 | 0.49 | 0.55 | 10.48 | cl | 0.78 | 0.2 | re | 0.84 | 47 | bar | -?Di | -no- | -?- | SB0 | |
| 1878 | 4928 | 031 | NGC4839 | 1877 | -1694 | 11.40 | 1.53 | 1.64 | 12.15 | ft | 0.58 | 0.7 | re | 0.45 | 04 | -no | exDi | -no- | exD | SA0 | * |
| 1931 | 5051 | 024 | | 2076 | -1806 | 13.63 | 0.47 | 0.50 | 9.09 | ft | 0.80 | 2.9 | ex | 0.90 | 60 | bar | -?Di | -no- | -?- | SB0 | |

Table 2a. Continued

| (1) | (2) | (3) | (4) | (5) | (6) | (7) | (8) | (9) | (10) | (11) | (12) | (13) | (14) | (15) | (16) | (17) | (18) | (19) | (20) | (21) | (22) |
|---|---|---|---|---|---|---|---|---|---|---|---|---|---|---|---|---|---|---|---|---|---|
| 1976 | 5160 | 245 |  | 2217 | 2348 | 14.29 | 0.48 | 0.72 | 9.77 | st | 0.36 | 12.1 | ex | 0.54 | 00 | -no | emDi | -no- | spH | SA0 |  |
| 2062 | 5364 | 035 | NGC4824 | 2543 | -1542 | 14.08 | 0.56 | 0.59 | 9.97 | ft | 0.83 | 1.9 | ex | 0.90 | 20 | -no | -?Di | -no | -?- | SA0 |  |
| 2074 | 5397 | 140 |  | 2589 | -98 | 13.88 | 0.80 | 0.82 | 10.98 | no | 0.97 | 0.4 | re | 0.95 | 90 | -no | -no- | -no- | -?- | unE | * |
| 2085 | 5428 |  |  | 2634 | -492 | 13.65 | 0.66 | 0.68 | 10.03 | no | 0.96 | -0.5 | re | 0.85 | 21 | -no | -no- | -no- | -?- | boE |  |
| 2118 | 5526 |  |  | 2778 | -1875 | 14.48 | 0.62 | 0.66 | 10.68 | ft | 0.88 | 2.5 | co | 0.66 | 80 | bar | -?Di | spiP? | -?- | SB0/a |  |
| 2134 | 5568 |  | NGC4816 | 2828 | -795 | 11.62 | 1.44 | 1.49 | 11.87 | no | 0.80 | 0.0 | re | 0.63 | 16 | -no | -no- | -no- | pec | unE | * |
| 2297 | 6043 |  |  | 3515 | -354 | 14.56 | 0.84 | 0.89 | 11.59 | st | 0.74 | 1.4 | ex | 0.74 | 00 | -no | -?Di | -no- | -?- | SA0 |  |
| 1205 | 3275 |  | RB039 | 76 | -419 | 15.98 | 0.41 | 0.45 | 11.13 | ft | 0.77 | -2.0 | ex | 0.86 | 40 | bar | -?Di | -no- | -?- | SB0 |  |
| 1322 | 3534 | 158 | RB007 | 315 | 11 | 15.58 | 0.43 | 0.51 | 10.81 | cl | 0.47 | 2.8 | ex | 0.51 | 05 | -no | miDi | -no- | thD | SA0 |  |
| 1328 | 3554 |  | RB271 | 332 | 374 | 15.91 | 0.70 | 0.72 | 12.52 | no | 0.85 | 1.5 | ex | 0.90 | 45 | bar | -?Di | -no- | -?- | SB0 |  |
| 1426 | 3794 | 134 | RB261 | 546 | -40 | 15.51 | 0.19 | 0.26 | 9.54 | ft | 0.68 | 2.1 | ex | 0.96 | 05 | -no | emDi | -no- | spH | diE/SA0 |  |
| 1533 | 4083 |  | RB198 | 856 | -516 | 15.95 | 0.48 | 0.57 | 11.46 | ft | 0.60 | 0.0 | co | 0.63 | 00 | -no | exDi | -no- | exD | SA0 |  |
| 1636 | 4341 | 073 | RB183 | 1142 | -748 | 15.21 | 0.33 | 0.45 | 9.95 | cl | 0.46 | 2.9 | ex | 0.62 | 02 | -no | exDi | -no- | exD | SA0 |  |
| 1647 | 4364 |  | RB182 | 1175 | -865 | 15.27 | 0.66 | 0.82 | 11.67 | – | – | – | – | — | — | — | — | — | spiP | S.. |  |
| 2022 | 5279 |  | NGC4827 | 2429 | -2840 | 11.90 | 1.11 | 1.17 | 10.53 | no | 0.78 | 0.8 | co | 0.85 | 15 | -no | emDi | -no- | -?- | diE |  |

Table 2b. Photometric parameters and classification of the CCD CFH subsample

| (1) | (2) | (3) | (4) | (5) | (6) | (7) | (8) | (9) | (10) | (11) | (12) | (13) | (14) | (15) | (16) | (17) | (18) | (19) | (20) | (21) | (22) |
|---|---|---|---|---|---|---|---|---|---|---|---|---|---|---|---|---|---|---|---|---|---|
| 0399 | 1245 | 095 |  | -2486 | -370 | 13.70 | 0.37 | 0.50 | 8.65 | st | 0.36 | 4.3 | ex | 0.40 | 00 | -no | emDi | -no- | spH | SA0 |  |
| 0565 | 1750 | 049 | NGC4926 | -1705 | -1245 | 12.36 | 0.95 | 1.01 | 10.21 | no | 0.81 | 0.8 | co | — | 09 | -no | -?Di | -no- | -?- | diE |  |
| 0661 | 2000 | 078 | NGC4923 | -1413 | -443 | 13.10 | 0.79 | 0.83 | 10.12 | no | 0.78 | 0.7 | ex | — | 35 | -no | -?Di | -no- | -?- | diE |  |
| 0733 | 2201 |  | RB129 | -1172 | -203 | 15.00 | 0.60 | 0.63 | 11.10 | no | 0.97 | 0.0 | re | 0.96 | — | -no | -no- | -no- | -?- | unE | * |
| 1039 | 2921 | 148 | NGC4889 | -304 | 22 | — | — | — | — | no | 0.65 | -0.7 | ex | — | 00 | -no | -no- | -no- | -?- | boE | * |
| 1065 | 2975 | 151 | NGC4886 | -254 | 61 | 13.55 | 0.73 | 0.73 | 10.32 | no | 0.97 | 0.0 | re | — | — | -no | -no- | -no- | -?- | unE | * |
| 1172 | 3201 | 124 | NGC4876 | 10 | -210 | — | — | — | — | ft | 0.64 | 0.6 | ex | 0.72 | 05 | -no | emDi | -no- | -?- | diE |  |
| 1331 | 3561 | 179 | NGC4865 | 335 | 409 | 13.00 | 0.67 | 0.83 | 9.57 | ft | 0.44 | 1.4 | ex | 0.54 | 00 | -no | emDi | -no- | -?- | diE | * |
| 1365 | 3639 | 133 | NGC4867 | 398 | 2 | — | — | — | — | ft | 0.65 | 1.0 | ex | 0.81 | 04 | -no | emDi | -no- | spH | diE |  |
| 1378 | 3664 | 159 | NGC4864 | 426 | 24 | 12.95 | 0.72 | 0.76 | 9.65 | no | 0.82 | 0.0 | re | — | 00 | -no | -no- | -no- | -?- | unE |  |
| 1821 | 4792 | 023 | NGC4842b | 1721 | -1743 | 14.45 | 0.41 | 0.50 | 9.58 | cl | 0.52 | 2.9 | ex | 0.65 | 05 | -no | emDi | -no- | spH | SA0 |  |
| 1823 | 4794 | 030 | NGC4842a | 1724 | -1712 | 13.42 | 0.60 | 0.62 | 9.50 | no | 0.89 | 0.0 | co | 0.96 | 07 | -no | -no- | -no- | -?- | unE |  |
| 1828 | 4806 | 239 | NGC4841b | 1729 | 1848 | 13.35 | 0.61 | 0.62 | 9.52 | no | 0.99 | 0.0 | co | — | — | -no | -no- | -no- | -?- | unE |  |
| 1831 | 4822 | 240 | NGC4841a | 1754 | 1828 | 12.40 | 0.84 | 0.88 | 9.72 | no | 0.85 | -1.0 | ex | — | — | -no | -no- | -no- | -?- | boE |  |
| 1878 | 4928 | 031 | NGC4839 | 1877 | -1694 | 11.60 | 1.05 | 1.14 | 9.94 | ft | 0.65 | 1.2 | ex | — | 03 | -no | -?Di | -no- | -?- | diE |  |
| 1883 | 4943 | 032 |  | 1896 | -1712 | 14.00 | 0.49 | 0.62 | 9.52 | no | 0.53 | -1.5 | co | — | 35 | -no | -no- | -no- | -?- | boE | * |
| 2109 | 5495 |  |  | 2727 | -778 | 13.20 | 0.64 | 0.69 | 9.50 | no | 0.82 | -0.3 | re | — | 13 | -no | -no- | -no- | -?- | unE | * |

Table 3. Photometric parameters and classification of the KP1608 sample

| (1) | (2) | (3) | (4) | (5) | (6) | (7) | (8) | (9) | (10) | (11) | (12) | (13) | (14) | (15) | (16) | (17) | (18) | (19) | (20) | (21) | (22) |
|---|---|---|---|---|---|---|---|---|---|---|---|---|---|---|---|---|---|---|---|---|---|
| 0884 | 2559 | 169 | IC4040 | -695 | 316 | 14.02 | 0.59 | 0.79 | 10.08 | — | — | — | — | — | — | — | — | — | — | S.. | * |
| 0893 | 2584 | 192 |  | -666 | -100 | 14.20 | 0.62 | 0.83 | 10.39 | st | 0.21 | 11 | ex | 0.29 | 02 | -no | emDi | -no- | thD | SA0 |  |
| 0917 | 2651 | 147 | RB100 | -571 | 6 | 14.50 | 0.70 | 0.82 | 11.11 | st | 0.37 | 4 | ex | 0.71 | 09 | -no | emDi | -no- | spH | SA0 |  |
| 0918 | 2654 | 119 | RB099 | -565 | -52 | 14.66 | 0.57 | 0.61 | 10.60 | st | 0.64 | 1 | co | 0.67 | 45 | bar | exDi | -no- | exD | SB0 |  |
| 0951 | 2727 | 170 | IC4026 | -490 | 276 | 13.96 | 0.79 | 0.84 | 11.00 | cl | 0.68 | 3 | ex | 0.71 | 50 | bar | exDi | -no- | exD | SB0 |  |
| 0969 | 2778 |  | RB094 | -445 | -122 | 15.20 | 0.57 | 0.72 | 11.16 | cl | 0.50 | 0 | co | 0.93 | 85 | bar | -?Di | spiP? | -?- | SB0/a | * |
| 0976 | 2794 | 120 | NGC4898b | -434 | -51 | — | — | — | — | — | — | — | — | — | — | — | — | — | — | ??? | * |
| 0977 | 2795 | 206 | NGC4895 | -433 | 834 | 12.50 | 1.00 | 1.13 | 10.60 | ft | 0.32 | 5 | ex | 0.36 | 03 | -no | emDi | -no- | thD | SA0 |  |
| 0980 | 2798 | 121 | NGC4898a | -429 | -54 | — | — | — | — | — | — | — | — | — | — | — | — | — | — | ??? | * |
| 0985 | 2805 | 171 | RB091 | -422 | 337 | 14.80 | 0.49 | 0.55 | 10.36 | cl | 0.52 | 0 | re | 0.64 | 27 | bar | emDi | -no- | spH | SB0 |  |
| 0990 | 2815 | 122 | NGC4894 | -415 | -11 | 14.38 | 0.62 | 0.72 | 10.56 | st | 0.38 | 7 | ex | 0.54 | 03 | -no | emDi | -no- | spH | SA0 |  |
| 0999 | 2839 | 172 | IC4021 | -392 | 255 | 14.33 | 0.53 | 0.55 | 10.09 | ft | 0.90 | 1 | ex | 0.78 | 20 | -no | -?Di | -no- | exD? | SA0 |  |
| 1011 | 2861 | 173 | RB087 | -367 | 378 | 14.52 | 0.54 | 0.61 | 10.34 | cl | 0.54 | 6 | ex | 0.65 | 09 | -no | emDi | spiP? | thD | SA0/a |  |
| 1030 | 2897 | 099 | RB083 | -330 | -384 | 15.31 | 0.61 | 0.62 | 11.44 | no | 0.80 | 0 | re | 0.92 | 20 | bar | -?Di | -no- | -?- | SB0 |  |
| 1033 | 2910 | 100 | RB082 | -317 | -376 | 14.89 | 0.52 | 0.58 | 10.59 | cl | 0.71 | -2 | ex | 0.90 | 15 | bar? | -?Di | spiP | -?- | SABa |  |
| 1035 | 2912 | 207 | RB167 | -316 | 719 | 14.29 | 0.60 | 0.67 | 10.41 | ft | 0.53 | 2 | ex | 0.57 | 07 | bar? | -?Di | -no- | -?- | SAB0 | * |
| 1039 | 2921 | 148 | NGC4889 | -304 | 22 | 10.80 | 1.48 | 1.58 | 11.32 | no | 0.59 | 2 | ex | 0.70 | 04 | -no | -no- | -no- | pec | boEp | * |
| 1040 | 2922 | 174 | IC4012 | -303 | 388 | 14.27 | 0.48 | 0.52 | 9.77 | no | 0.75 | 2 | ex | 0.89 | 15 | -no | emDi | -no- | -?- | diE |  |
| 1046 | 2940 | 150 | IC4011 | -280 | 120 | 14.38 | 0.66 | 0.67 | 10.76 | no | 0.90 | -1 | ex | 0.99 | 00 | -no | -no- | -no- | -?- | boE |  |
| 1050 | 2946 | 149 | RB077 | -277 | 28 | 14.80 | 0.71 | 0.88 | 11.42 | st | 0.32 | 7 | ex | 0.47 | 01 | -no | emDi | -no- | spH | SA0 | * |
| 1057 | 2956 | 084 |  | -268 | -587 | 14.23 | 0.62 | 0.78 | 10.42 | st | 0.32 | 5 | ex | 0.39 | 01 | -no | miDi | -no- | thD | SA0 |  |
| 1059 | 2960 |  | RB074 | -267 | 194 | 15.00 | 0.65 | 0.72 | 11.33 | cl | 0.59 | 4 | ex | 0.49 | 02 | -no | exDi | -no- | exD | SA0 |  |
| 1065 | 2975 | 151 | NGC4886 | -254 | 61 | 13.50 | 0.88 | 0.90 | 11.02 | no | 0.97 | 1 | re | 0.87 | — | -no | -no- | -no- | -?- | unE |  |
| 1104 | 3055 | 217 | NGC4881 | -168 | 994 | 12.80 | 1.11 | 1.11 | 11.46 | no | 0.98 | 0 | re | 0.97 | — | -no | -no- | -no- | -?- | unE |  |
| 1106 | 3068 | 123 | RB064 | -151 | -147 | 14.68 | 0.72 | 0.85 | 11.36 | — | — | — | — | — | — | -no | — | SpiP | — | Sa |  |
| 1109 | 3073 | 175 | NGC4883 | -144 | 230 | 13.72 | 0.69 | 0.74 | 10.29 | cl | 0.70 | 0 | co | 0.84 | 32 | bar? | -?Di | -no- | pec | SAB0 | * |
| 1115 | 3084 | 193 | RB155 | -132 | 568 | 14.77 | 0.49 | 0.53 | 10.34 | ft | 0.80 | 2 | ex | 0.95 | 30 | bar? | -?Di | -no- | -?- | SAB0 | * |
| 1157 | 3170 | 152 | IC3998 | -20 | 12 | 14.07 | 0.73 | 0.77 | 10.84 | st | 0.64 | 2 | ex | 0.75 | 60 | bar | emDi | -no- | thD | SB0 |  |
| 1164 | 3178 | 101 | RB049 | -11 | -410 | 14.26 | 0.64 | 0.74 | 10.58 | cl | 0.61 | 9 | ex | 0.33 | 03 | -no | exDi | -no- | exD | SA0 |  |
| 1172 | 3201 | 124 | NGC4876 | 10 | -210 | 13.85 | 0.62 | 0.67 | 10.03 | ft | 0.66 | 1 | co | 0.74 | 06 | -no | -?Di | -no- | -?- | diE/SA0 |  |
| 1175 | 3206 | 126 | RB046 | 14 | -43 | 14.98 | 0.49 | 0.61 | 10.51 | cl | 0.50 | 2 | co | 0.55 | 03 | -no | emDi | -no- | thD | SA0 |  |
| 1177 | 3213 | 153 | RB045 | 19 | 87 | 14.63 | 0.52 | 0.54 | 10.34 | ft | 0.90 | 2 | ex | 0.92 | 35 | -no | -?Di | -no- | -?- | diE | * |
| 1181 | 3222 | 125 | RB043 | 38 | -166 | 14.94 | 0.34 | 0.35 | 9.71 | no | 0.78 | -1 | ex | 0.95 | 20 | -no | -no- | -no- | pec | boE | * |
| 1194 | 3254 | 127 | RB042 | 65 | -8 | 15.27 | 0.48 | 0.51 | 10.76 | st | 0.53 | 3 | ex | 0.60 | 03 | -no | emDi | -no- | thD | SA0 |  |
| 1201 | 3269 | 128 | RB040 | 74 | -60 | 15.11 | 0.41 | 0.51 | 10.24 | st | 0.45 | 3 | ex | 0.48 | 01 | -no | emDi | -no- | thD | SA0 |  |
| 1210 | 3291 | 154 | RB038 | 92 | 61 | 15.00 | 0.72 | 0.83 | 11.69 | cl | 0.62 | 2 | co | 0.73 | 05 | -no | emDi | -no- | spH | SA0 |  |
| 1214 | 3296 | 104 | NGC4875 | 98 | -229 | 14.23 | 0.51 | 0.55 | 9.87 | cl | 0.68 | 2 | ex | 0.77 | 10 | -no | emDi | -no- | -?- | SA0 |  |
| 1233 | 3329 | 129 | NGC4874 | 124 | -41 | 10.90 | 1.71 | 1.74 | 12.57 | no | 0.87 | 0 | re | 0.88 | 30 | -no | -no- | -no- | pec | unE | * |
| 1245 | 3352 | 130 | NGC4872 | 147 | -86 | 13.88 | 0.65 | 0.67 | 10.20 | st | 0.57 | 3 | ex | 0.60 | 35 | bar | -?Di | -no- | -?- | SB0 |  |
| 1250 | 3367 | 155 | NGC4873 | 166 | 47 | 13.71 | 0.73 | 0.77 | 10.47 | cl | 0.68 | 2 | co | 0.73 | 00 | -no | emDi | -no- | -?- | SA0 |  |
| 1258 | 3390 | 176 | RB026 | 183 | 275 | 14.29 | 0.51 | 0.62 | 9.92 | cl | 0.43 | 4 | ex | 0.60 | 03 | -no | emDi | -no- | spH | SA0 |  |
| 1264 | 3400 | 103 | IC3973 | 191 | -310 | 13.72 | 0.70 | 0.76 | 10.32 | — | — | — | — | — | — | — | — | — | — | S.. | * |
| 1266 | 3403 | 087 | RB234 | 194 | -644 | 15.15 | 0.44 | 0.45 | 10.43 | no | 0.88 | 1 | co | 0.95 | 20 | -no | -?Di | -no- | -?- | diE |  |
| 1270 | 3414 | 131 | NGC4871 | 202 | -52 | 13.70 | 0.75 | 0.84 | 10.54 | st | 0.53 | 3 | co | 0.79 | 13 | bar? | -?Di | SpiP | -?- | SAB0/a |  |
| 1274 | 3423 | 088 | IC3976 | 211 | -434 | 14.08 | 0.54 | 0.67 | 9.88 | cl | 0.42 | 3 | ex | 0.57 | 00 | -no | emDi | -no- | thD | SA0 |  |

| (1) | (2) | (3) | (4) | (5) | (6) | (7) | (8) | (9) | (10) | (11) | (12) | (13) | (14) | (15) | (16) | (17) | (18) | (19) | (20) | (21) | (22) |
|---|---|---|---|---|---|---|---|---|---|---|---|---|---|---|---|---|---|---|---|---|---|
| 1277 | 3433 | 177 | RB022 | 219 | 252 | 15.00 | 0.46 | 0.53 | 10.40 | ft | 0.58 | 0 | co | 0.59 | 18 | -no | -?Di | -no- | -?- | SA0/diE | * |
| 1280 | 3439 | 178 | RB021 | 221 | 413 | 15.26 | 0.58 | 0.60 | 11.28 | cl | 0.88 | 0 | co | 0.83 | 30 | bar | -?Di | -no- | -?- | SB0 | |
| 1291 | 3471 | 156 | RB018 | 247 | 101 | 15.25 | 0.42 | 0.46 | 10.45 | ft | 0.68 | 2 | ex | 0.81 | 20 | bar | -?Di | -no- | -?- | SB0 | |
| 1300 | 3484 | 157 | RB014 | 262 | 9 | 14.70 | 0.56 | 0.61 | 10.60 | cl | 0.73 | 0 | re | 0.50 | 11 | -no | exDi | spiP | exD | SAa | |
| 1302 | 3487 | 132 | RB013 | 263 | -9 | 14.80 | 0.61 | 0.67 | 10.94 | cl | 0.52 | 4 | ex | 0.72 | 10 | bar? | -?Di | -no- | -?- | SAB0 | |
| 1304 | 3493 | 067 | RB230 | 270 | -833 | 14.52 | 0.48 | 0.54 | 10.02 | ft | 0.44 | 5 | ex | 0.52 | 10 | -no | miDi | -no- | thD | SA0 | |
| 1311 | 3509 | 236 | | 288 | 1864 | 15.10 | 0.48 | 0.60 | 10.61 | st | 0.59 | 5 | ex | 0.66 | 10 | -no | -?Di | spiP | -?- | SA0/a | |
| 1312 | 3510 | 105 | NGC4869 | 289 | -214 | 13.14 | 0.79 | 0.84 | 10.20 | no | 0.84 | 0 | co | 0.94 | 10 | -no | -no- | -no- | -?- | unE | |
| 1318 | 3522 | 106 | RB008 | 300 | -266 | 14.75 | 0.47 | 0.51 | 10.18 | cl | 0.73 | 0 | co | 0.85 | 10 | bar | -?Di | -no- | -?- | SB0 | |
| 1327 | 3553 | 208 | RB136 | 331 | 819 | 15.00 | 0.62 | 0.70 | 11.20 | cl | 0.56 | 1 | ex | 0.63 | 09 | -no | -?Di | -no- | -?- | SA0 | |
| 1329 | 3557 | 107 | RB006 | 334 | -304 | 14.74 | 0.52 | 0.64 | 10.45 | cl | 0.57 | 1 | co | 0.84 | 05 | -no | emDi | -no- | spH | SA0 | |
| 1331 | 3561 | 179 | NGC4865 | 335 | 409 | 12.88 | 0.71 | 0.81 | 9.53 | ft | 0.44 | 1 | ex | 0.53 | 00 | -no | emDi | -no- | spH? | diE/SA0 | |
| 1365 | 3639 | 133 | NGC4867 | 398 | 2 | 13.80 | 0.59 | 0.65 | 9.84 | no | 0.71 | 1 | co | 0.73 | 08 | -no | -?Di | -no- | -?- | diE | |
| 1373 | 3656 | 180 | RB268 | 414 | 382 | 13.81 | 1.08 | 1.09 | 12.31 | no | 0.88 | 0 | re | 0.96 | 20 | -no | -no- | -no- | -?- | unE/SA0 | * |
| 1374 | 3660 | 068 | IC3963 | 422 | -704 | 13.98 | 0.77 | 0.86 | 10.93 | st | 0.49 | 10 | ex | 0.80 | 05 | -no | emDi | -no- | spH | SA0 | |
| 1378 | 3664 | 159 | NGC4864 | 426 | 24 | 13.05 | 0.73 | 0.81 | 9.82 | no | 0.75 | 0 | re | 0.96 | 08 | -no | -no- | -no- | -?- | unE | * |
| 1401 | 3730 | 069 | IC3959 | 492 | -670 | 13.56 | 0.71 | 0.74 | 10.22 | no | 0.85 | -1 | co | 0.90 | 20 | -no | -no- | -no- | -?- | unE | |
| 1402 | 3733 | 109 | IC3960 | 496 | -415 | 14.07 | 0.64 | 0.69 | 10.38 | cl | 0.81 | 1 | ex | 0.94 | 70 | bar | -?Di | -no- | -?- | SB0 | |
| 1406 | 3739 | 070 | IC3957 | 502 | -729 | 14.18 | 0.64 | 0.66 | 10.50 | no | 0.90 | 2 | ex | 0.92 | 15 | -no | -?Di | -no- | -?- | diE | |
| 1414 | 3761 | 160 | IC3955 | 519 | 96 | 13.75 | 0.77 | 0.82 | 10.70 | st | 0.54 | 5 | ex | 0.73 | 30 | bar? | -?Di | spiP | -?- | SAB0/a | * |
| 1423 | 3782 | | RB262 | 537 | -213 | 14.81 | 0.51 | 0.60 | 10.43 | st | 0.46 | 5 | ex | 0.59 | 05 | -no | emDi | -no- | spH | SA0 | |
| 1425 | 3792 | 194 | NGC4860 | 545 | 552 | 12.80 | 0.89 | 0.93 | 10.35 | no | 0.82 | 1 | co | 0.89 | 08 | -no | -?Di | -no- | -?- | diE | |
| 1431 | 3816 | 195 | NGC4858 | 570 | 523 | 14.45 | 0.59 | 0.68 | 10.62 | – | — | – | – | – | – | — | spiP | — | S.. | | |
| 1432 | 3818 | 218 | | 573 | 919 | 13.53 | 0.72 | 0.78 | 10.24 | cl | 0.42 | 3 | ex | 0.43 | 13 | bar? | exDi | -no- | exD | SAB0 | |
| 1444 | 3851 | 135 | RB260 | 599 | -9 | 15.18 | 0.50 | 0.56 | 10.76 | no | 0.78 | 2 | ex | 0.74 | 08 | -no | -?Di | -no- | -?- | diE | |
| 1454 | 3882 | 071 | RB214 | 633 | -665 | 14.99 | 0.61 | 0.77 | 11.13 | st | 0.33 | 5 | ex | 0.36 | 01 | -no | emDi | -no- | thD | SA0 | |
| 1456 | 3892 | 237 | | 645 | 1776 | 14.93 | 0.67 | 0.72 | 11.37 | cl | 0.63 | 2 | ex | 0.66 | 30 | bar | exDi | -no- | exD | SB0/a | |
| 1458 | 3896 | | IC3949 | 646 | -491 | 13.17 | 0.85 | 1.20 | 10.52 | st | 0.14 | 4 | ex | 0.20 | 02 | -no | exDi | spiP | exD | S.. | * |
| 1466 | 3914 | 136 | RB257 | 661 | -19 | 14.88 | 0.38 | 0.45 | 9.85 | ft | 0.72 | 2 | ex | 0.95 | 00 | -no | emDi | -no- | -?- | diE/SA0 | |
| 1476 | 3935 | 196 | | 684 | 561 | 14.65 | 0.54 | 0.60 | 10.45 | ft | 0.67 | 3 | co | 0.67 | 25 | -no | -?Di | spiP | -?- | SA0/a | |
| 1480 | 3943 | 090 | RB209 | 692 | -564 | 15.05 | 0.58 | 0.71 | 11.06 | cl | 0.44 | 2 | ex | 0.72 | 04 | -no | emDi | -no- | spH | SA0 | |
| 1485 | 3958 | 72 | IC3947 | 707 | -666 | 14.17 | 0.54 | 0.63 | 9.98 | cl | 0.65 | 3 | ex | 0.80 | 05 | -no | emDi | -no- | spH | SA0 | |
| 1489 | 3972 | 181 | RB252 | 721 | 410 | 14.56 | 0.58 | 0.64 | 10.56 | st | 0.28 | 15 | ex | 0.32 | 05 | -no | exDi | -no- | exD | SA0 | * |
| 1501 | 3997 | 091 | IC3946 | 751 | -575 | 13.43 | 0.66 | 0.72 | 9.84 | cl | 0.41 | 5 | ex | 0.45 | 09 | bar? | miDi | spiP? | thD | SAB0/a | |
| 1507 | 4017 | 058 | NGC4854 | 769 | -1064 | 13.26 | 1.07 | 1.10 | 11.71 | ft | 0.70 | 0 | co | 0.73 | 35 | bar | -?Di | -no- | -?- | SB0 | |
| 1553 | 4130 | 197 | IC3943 | 909 | 517 | 13.60 | 0.71 | 0.86 | 10.26 | – | — | – | – | – | – | — | spiP | — | Sa | | |
| 1581 | 4200 | 182 | RB243 | 971 | 288 | 15.17 | 0.34 | 0.38 | 10.05 | no | 0.85 | 1 | ex | 0.90 | 25 | Bar? | -?Di | -no- | -?- | SAB0 | |
| 1594 | 4230 | 161 | RB241 | 992 | 162 | 13.31 | 0.89 | 0.92 | 10.87 | no | 0.85 | 0 | co | 0.89 | 15 | -no | -no- | -no- | — | unE | |
| 1597 | 4235 | 225 | | 998 | 1193 | 15.42 | 0.47 | 0.58 | 10.88 | cl | 0.49 | 3 | ex | 0.63 | 03 | -no | emDi | -no- | spH | SA0 | |
| 1601 | 4242 | 243 | | 1007 | 2006 | 14.94 | 0.51 | 0.54 | 10.58 | cl | 0.60 | 2 | ex | 0.60 | 10 | -no | exDi | spiP | exD | SA0/a | |
| 1620 | 4308 | 198 | | 1098 | 657 | 14.71 | 0.61 | 0.65 | 10.89 | st | 0.53 | 3 | ex | 0.70 | 12 | -no | emDi | -no- | -?- | SA0 | * |
| 1623 | 4313 | 199 | NGC4851 | 1103 | 645 | 14.72 | 0.61 | 0.71 | 10.85 | st | 0.58 | 6 | ex | 0.61 | 45 | bar | -?Di | spiP | -?- | SBa | * |
| 1625 | 4315 | 137 | NGC4850 | 1104 | -6 | 13.60 | 0.70 | 0.75 | 10.20 | cl | 0.77 | 3 | ex | 0.99 | 28 | Bar? | -?Di | -no- | -?- | SAB0 | |
| 1655 | 4379 | 200 | | 1193 | 562 | 14.23 | 0.42 | 0.52 | 9.44 | cl | 0.41 | 10 | ex | 0.56 | 01 | -no | emDi | -no- | spH | SA0 | |
| 1662 | 4391 | 209 | | 1207 | 767 | 14.22 | 0.85 | 1.02 | 11.57 | cl | 0.40 | 2 | ex | 0.43 | 07 | -no | exDi | -no- | exD | SA0 | |

**Table 3.** Continued

| (1) | (2) | (3) | (4) | (5) | (6) | (7) | (8) | (9) | (10) | (11) | (12) | (13) | (14) | (15) | (16) | (17) | (18) | (19) | (20) | (21) | (22) |
|---|---|---|---|---|---|---|---|---|---|---|---|---|---|---|---|---|---|---|---|---|---|
| 1688 | 4471 | 220 | NGC4848 | 1311 | 976 | 12.76 | 0.86 | 1.04 | 10.13 | – | — | – | – | — | – | — | — | — | — | S.. | |
| 1696 | 4499 | 092 | | 1350 | -556 | 14.34 | 0.50 | 0.55 | 9.96 | cl | 0.59 | 6 | ex | 0.88 | 10 | -no | emDi | -no- | spH | SA0 | |
| 1700 | 4503 | 138 | | 1351 | 3 | 14.97 | 0.70 | 0.83 | 11.54 | – | — | – | – | — | – | — | — | — | — | S.. | |
| 1724 | 4555 | 183 | | 1420 | 331 | — | — | — | — | – | — | – | – | — | – | — | — | — | — | ??? | * |
| 1739 | 4588 | 238 | | 1464 | 1909 | 14.68 | 0.48 | 0.50 | 10.17 | ft | 0.73 | 2 | ex | 0.70 | 06 | -no | emDi | -no- | -?- | diE | * |
| 1758 | 4626 | 110 | | 1519 | -323 | 14.72 | 0.56 | 0.59 | 10.60 | cl | 0.65 | 3 | ex | 0.90 | 40 | -no | -?Di | spiP? | -?- | SA0/a | * |
| 1768 | 4648 | 210 | | 1540 | 760 | 14.25 | 0.54 | 0.69 | 10.03 | cl | 0.71 | 3 | ex | 0.90 | 46 | bar | -?Di | -no- | -?- | SB0 | |
| 1770 | 4653 | 111 | | 1554 | -311 | 13.88 | 0.74 | 0.77 | 10.69 | cl | 0.70 | 0 | re | 0.66 | 09 | -no | exDi | -no- | exD | SA0 | |
| 1777 | 4664 | 093 | | 1564 | -490 | 14.20 | 0.51 | 0.59 | 9.85 | st | 0.28 | 13 | ex | 0.36 | 04 | -no | miDi | -no- | thD | SA0 | |
| 1781 | 4679 | 075 | | 1582 | -764 | 14.20 | 0.74 | 0.97 | 10.98 | st | 0.20 | 7 | ex | 0.26 | 04 | -no | emDi | -no- | thD | SA0 | |
| 1844 | 4849 | 211 | | 1786 | 789 | 13.55 | 0.85 | 1.00 | 10.92 | st | 0.34 | 7 | ex | 0.50 | 07 | -no | emDi | -no- | spH | SA0 | * |
| 1852 | 4866 | 212 | | 1807 | 748 | 14.20 | 0.72 | 0.74 | 10.89 | st | 0.48 | 3 | ex | 0.59 | 22 | bar | miDi | -no- | thD | SAB0 | |
| 1879 | 4933 | 076 | | 1882 | -729 | 14.55 | 0.76 | 0.89 | 11.43 | st | 0.40 | 10 | ex | 0.85 | 14 | -no | -?Di | spiP | -?- | SA0/a | |
| 1885 | 4945 | 112 | | 1902 | -317 | 14.91 | 0.51 | 0.53 | 10.58 | ft | 0.71 | 3 | ex | 0.73 | 31 | bar | -?Di | -no- | -?- | SB0 | |
| 1900 | 4974 | 094 | | 1956 | -567 | 14.85 | 0.50 | 0.54 | 10.44 | cl | 0.55 | 1 | ex | 0.63 | 03 | -no | emDi | -no- | thD | SA0 | |
| 1986 | 5191 | 139 | | 2281 | -139 | 14.77 | 0.34 | 0.42 | 9.57 | ft | 0.59 | 4 | ex | 0.73 | 07 | -no | emDi | -no- | spH | diE/SA0 | |
| 2020 | 5272 | 163 | NGC4828 | 2413 | 191 | 13.41 | 0.91 | 0.91 | 11.07 | no | 0.93 | 0 | re | 0.88 | – | -no | -no- | -no- | pec | unE | * |
| 1076 | 2989 | 216 | RB160 | -237 | 971 | 15.55 | 0.58 | 0.60 | 11.56 | ft | 0.59 | 0 | re | 0.65 | 25 | -no | -?Di | spiP | -?- | Sa | |
| 1687 | 4470 | | | 1310 | 1444 | 15.30 | 0.72 | 0.81 | 12.00 | cl | 0.57 | 4 | ex | 0.56 | 06 | -no | exDi | spiP | exD | SA0/a | * |
| 1778 | 4666 | 201 | | 1566 | 618 | 15.33 | 0.62 | 0.77 | 11.52 | cl | 0.37 | 3 | ex | 0.39 | 10 | bar? | exDi | spiP? | exD | SAB0/a | * |
| 1817 | 4779 | 184 | | 1703 | 389 | 15.12 | 0.70 | 0.76 | 11.71 | st | 0.62 | 6 | ex | 0.62 | 16 | -no | exDi | spiP | exD | SA0/a | |
| 2015 | 5256 | 185 | | 2374 | 301 | 15.25 | 0.64 | 0.64 | 11.44 | ft | 0.85 | 0 | co | 0.80 | 16 | bar | -?Di | -no- | -?- | SB0 | |
| 2034 | 5304 | 186 | | 2465 | 410 | 15.10 | 0.78 | 0.82 | 12.11 | st | 0.80 | 3 | ex | 0.79 | 25 | bar | -?Di | -no- | -?- | SAB0 | * |

Table 4. Photometric parameters and classification of the FW sample

| (1) | (2) | (3) | (4) | (5) | (6) | (7) | (8) | (9) | (10) | (11) | (12) | (13) | (14) | (15) | (16) | (17) | (18) | (19) | (20) | (21) | (22) |
|---|---|---|---|---|---|---|---|---|---|---|---|---|---|---|---|---|---|---|---|---|---|
| 12 44 15 | 27 14 16 | | FW1 | | | 15.04 | 0.66 | 0.76 | 11.46 | no | 0.65 | 0 | er | 0.64 | 4 | -no | -no- | -no- | -?- | unE | |
| 12 44 25 | 26 58 56 | | anon01 | | | 15.40 | 0.36 | 0.48 | 10.32 | ft | 0.52 | 1 | co | 0.56 | 2 | -no | exDi | -no- | exD | SA0 | |
| 12 44 39 | 26 58 54 | | D15 | | | 14.90 | 0.44 | 0.57 | 10.19 | cl | 0.39 | 2 | co | 0.39 | 0 | -no | emDi | -no- | thD | SA0 | |
| 12 44 45 | 26 59 11 | | D16 | | | 15.08 | — | — | — | – | 0.15 | 3 | ex | 0.17 | 0 | -no | exDi | -no- | miDi | S.. | * |
| 12 45 07 | 27 38 08 | | anon02 | | | 15.94 | 0.23 | 0.32 | 10.18 | cl | 0.54 | 2 | ex | 0.64 | 0 | -no | -?Di | -no- | spH | SA0 | |
| 12 45 15 | 27 15 16 | | D1 | | | 12.98 | 0.77 | 0.93 | 9.91 | st | 0.25 | 10 | ex | 0.28 | 0 | -no | miDi | -no- | thD | SA0 | |
| 12 45 15 | 27 33 57 | | D4 | | | 15.53 | 0.31 | 0.50 | 10.29 | st | 0.33 | 11 | ex | 0.37 | 0 | -no | emDi | -no- | spH | SA0 | |
| 12 45 23 | 27 16 53 | | FW10 | | | 13.10 | 0.73 | 0.99 | 9.83 | st | 0.24 | 9 | ex | 0.28 | 0 | -no | -?Di | -no- | -?- | SA0 | |
| 12 45 29 | 27 29 46 | | NGC4692 | | | 12.30 | 1.10 | 1.18 | 10.91 | no | 0.79 | 1 | ex | 0.90 | 40 | -no | -?Di | -no- | -?- | diE | |
| 12 45 36 | 27 33 14 | | D5 | | | 15.28 | 0.46 | 0.69 | 10.66 | cl | 0.23 | 10 | ex | 0.39 | 2 | -no | emDi | -no- | spH | SA0 | |
| 12 45 38 | 26 55 04 | | FW2 | | | 13.89 | 0.70 | 0.74 | 10.48 | no | 0.85 | -1 | co | 0.74 | 0 | -no | -no- | -no- | -?- | boE | * |
| 12 45 42 | 27 12 42 | | D2 | | | 14.26 | 0.51 | 0.71 | 9.91 | st | 0.21 | 10 | ex | 0.27 | 0 | -no | emDi | -no- | spH | SA0 | |
| 12 45 54 | 27 05 22 | | FW11 | | | 16.05 | 0.32 | 0.35 | 10.73 | no | 0.79 | -2 | ex | 0.86 | 30 | -no | -no- | -no- | -?- | boE | |
| 12 45 57 | 27 07 49 | | D3 | | | 13.31 | 0.69 | 0.85 | 9.84 | st | 0.31 | 6 | ex | 0.41 | 6 | -no | emDi | spiP? | spH | SA0/a | * |
| 12 46 20 | 27 01 21 | | anon03 | | | 15.79 | — | — | — | – | — | – | – | — | – | -no | — | — | — | S.. | * |
| 12 46 22 | 27 18 26 | | D8 | | | 14.32 | 0.75 | 0.95 | 11.16 | st | 0.30 | 3 | co | 0.34 | 0 | -no | emDi | -no- | thD | SA0 | |
| 12 46 34 | 27 27 05 | | NGC4702 | | | 15.35 | — | — | — | – | 0.55 | -4 | er | 0.58 | 40 | — | — | — | — | Irr | |
| 12 46 36 | 27 09 56 | | D11 | | | 14.15 | 0.55 | 0.68 | 9.97 | st | 0.29 | 11 | ex | 0.35 | 5 | -no | emDi | -no- | thD | SA0 | * |
| 12 46 42 | 27 38 29 | | D9 | | | 14.20 | 0.49 | 0.57 | 9.73 | st | 0.33 | 7 | ex | 0.48 | 0 | -no | miDi | -no- | spH | SA0 | |
| 12 46 59 | 27 03 32 | | D7 | | | 15.51 | 0.39 | 0.65 | 10.56 | st | 0.20 | 8 | ex | 0.23 | 0 | -no | emDi | -no- | thD | SA0 | |
| 12 47 01 | 27 32 24 | | FW4 | | | 15.07 | 0.62 | 0.67 | 11.25 | no | 0.78 | 2 | co | 0.96 | 0 | -no | -?Di | -no- | -?- | diE | |
| 12 47 09 | 27 28 11 | | D10 | | | 14.73 | 0.55 | 0.75 | 10.55 | st | 0.26 | 9 | ex | 0.36 | 0 | -no | emDi | -no- | thD | SA0 | * |
| 12 47 16 | 27 09 52 | | FW9 | | | 12.97 | 0.85 | 0.91 | 10.32 | no | 0.76 | 0 | er | 0.69 | 5 | -no | -no- | -no- | -?- | unE | * |
| 12 47 25 | 27 10 36 | | D12 | | | 14.95 | 0.52 | 0.65 | 10.67 | st | 0.31 | 4 | ex | 0.37 | 5 | -no | miDi | -no- | thD | SA0 | * |
| 12 47 28 | 27 08 49 | | FW6 | | | 15.95 | 0.30 | 0.29 | 10.57 | no | 0.81 | – | – | 0.81 | 60 | bar | -no- | -no- | -?- | SB0 | |
| 12 47 33 | 27 25 29 | | D13 | | | 15.56 | 0.40 | 0.51 | 10.66 | ft | 0.59 | 2 | ex | 0.45 | 0 | -no | exDi | -no- | -?- | SA0/diE | |
| 12 44 13 | 27 14 44 | | supp01 | | | 16.46 | 0.14 | 0.21 | 10.26 | no | 0.65 | 2 | ex | 0.70 | 5 | -no | -?Di | -no- | -?- | diE/SA0 | * |
| 12 44 23 | 27 04 06 | | supp02 | | | 16.12 | 0.37 | 0.43 | 11.07 | no | 0.58 | 2 | ex | 0.92 | 0 | bar? | -?Di | -no- | -?- | diE/SAB0 | |
| 12 44 54 | 27 44 10 | | supp03 | | | 16.41 | 0.33 | 0.36 | 11.17 | no | 0.83 | 1 | ex | 0.65 | 10 | -no | -?Di | -no- | -?- | diE/SA0 | |
| 12 45 17 | 27 31 13 | | supp04 | | | 16.66 | 0.19 | 0.59 | 10.70 | ft | 0.40 | 0 | ex | 0.56 | 5 | -no | emDi | -no- | spH | SA0 | |
| 12 45 26 | 27 32 22 | | supp05 | | | 16.66 | 0.20 | 0.26 | 10.76 | no | 0.72 | 1 | ex | 0.80 | 20 | -no | -?Di | -no- | -?- | diE | |
| 12 45 36 | 27 33 51 | | supp06 | | | 16.65 | — | — | — | st | 0.34 | 5 | ex | 0.63 | 5 | -no | — | spiP | — | S.. | * |
| 12 45 41 | 27 27 41 | | D14 | | | 17.18 | — | — | — | st | 0.27 | 3 | ex | 0.31 | 5 | -no | emDi | — | thD | S.. | * |
| 12 46 10 | 27 38 20 | | supp07 | | | 16.85 | 0.42 | 0.58 | 12.06 | no | 0.51 | 0 | ex | 0.79 | 30 | bar | -?Di | -no- | -?- | SB0 | |
| 12 46 20 | 26 01 43 | | supp08 | | | 16.61 | — | — | — | – | — | – | – | — | – | -no | — | — | — | S.. | * |
| 12 46 26 | 26 53 55 | | D6 | | | 16.36 | — | — | — | st | 0.24 | 5 | ex | 0.30 | 3 | -no | miDi | spiP? | spH | Sa | * |
| 12 46 26 | 26 55 55 | | supp09 | | | 16.61 | — | — | — | st | 0.19 | 10 | ex | 0.19 | 7 | -no | emDi | — | spH | S.. | |
| 12 46 33 | 27 38 51 | | supp10 | | | 16.22 | — | — | — | st | 0.22 | – | – | 0.39 | 5 | -no | exDi | — | spH | Irr | |
| 12 46 46 | 27 39 27 | | supp11 | | | 17.21 | — | — | — | st | 0.40 | 3 | ex | 0.50 | 20 | bar | -?Di | spiP | — | S.. | |
| 12 46 51 | 27 29 34 | | supp12 | | | 16.42 | 0.27 | 0.39 | 10.88 | st | 0.34 | 7 | ex | 0.37 | 5 | -no | emDi | -no- | thD | SA0 | * |
| 12 46 57 | 27 00 59 | | supp13 | | | 16.40 | 0.47 | 0.61 | 11.86 | cl | 0.37 | 4 | ex | 0.37 | 5 | -no | emDi | -no- | thD | SA0 | |
| 12 46 59 | 27 06 12 | | FW7 | | | 16.25 | 0.27 | 0.31 | 10.69 | no | 0.78 | 0 | co | 0.95 | 50 | bar? | -?Di | -no- | -?- | SAB0 | |
| 12 47 21 | 27 00 22 | | FW8 | | | 16.80 | 0.40 | 0.45 | 11.92 | no | 0.55 | 0 | co | 0.81 | 40 | bar | -no- | -no- | -?- | Irr | |
| 12 47 41 | 27 06 02 | | supp14 | | | 16.13 | 0.26 | 0.28 | 10.52 | cl | 0.53 | 3 | ex | 0.62 | 5 | -no | miDi | -no- | spH | SA0 | |
| 12 47 43 | 27 47 34 | | supp15 | | | 16.67 | 0.47 | 0.55 | 12.14 | no | 0.74 | 2 | ex | 0.85 | 50 | bar | -?Di | -no- | -?- | SB0 | |
| 12 47 49 | 27 17 43 | | supp16 | | | 16.97 | 0.31 | 0.40 | 11.63 | ft | 0.63 | 2 | co | — | 0 | -no | -?Di | -no- | -?- | diE/SA0 | |
| 12 47 49 | 27 19 57 | | FW5 | | | 16.60 | 0.37 | 0.36 | 11.55 | no | 0.94 | 0 | ex | 0.95 | 0 | -no | -no- | -no- | -?- | unE | |